\documentclass[a4paper,11pt,bookmarks=true]{article}
\pdfoutput=1
\usepackage{jheppub}

\usepackage[colorlinks=true,linkcolor=blue,citecolor=blue, urlcolor=blue]{hyperref}
\usepackage[utf8]{inputenc}
\usepackage{amssymb}
\usepackage{amsmath}\usepackage[normalem]{ulem}
\usepackage{color}
\usepackage{xspace}
\usepackage{graphicx}

\newcommand{\snn}{\ensuremath{ \sqrt{s_{\rm NN} } } }

\newcommand{\pp}{\ensuremath{\rm pp}}
\newcommand{\der}{\text{d}}
\newcommand{\ud}{\mathrm{d}}
\newcommand{\dNdy}{\ensuremath{\der N / \der y}}

\newcommand{\muB}{\ensuremath{\mu_{\rm B}} }

\newcommand{\jpsi}{\ensuremath{\text{J}/\psi}\xspace}
\newcommand{\LC}{\ensuremath{\Lambda_{\cQ}}\xspace}

\newcommand{\Dzero}{\ensuremath{\text{D}^{\text{0}}}\xspace}

\newcommand{\pt}{\ensuremath{p_{\text{T}}}\xspace}
\newcommand{\pT}{\ensuremath{p_{\text{T}}}\xspace}
\newcommand{\cQ}{\text{c}}

\newcommand{\ccBar}{\ensuremath{\text{c}\overline{\text{c}}}\xspace}
\newcommand{\cme}[1]{\ensuremath{\snn = #1{\rm\,TeV}}\xspace}

\usepackage{graphicx}

\preprint{CERN-TH-2021-057}
\title{The multiple-charm hierarchy in the statistical hadronization model}

  \author[a]{Anton Andronic,}
    \affiliation[a]{Institut f\"{u}r Kernphysik, Westf\"{a}lische Wilhelms-Universit\"{a}t M\"{u}nster, M\"{u}nster, Germany}
  \author[b,c,d]{Peter Braun-Munzinger,}
    \affiliation[b]{Research Division and ExtreMe Matter Institute EMMI, GSI Helmholtzzentrum f\"{u}r Schwerionenforschung GmbH, Darmstadt, Germany}
  \affiliation[c]{Physikalisches Institut, Ruprecht-Karls-Universit\"{a}t Heidelberg, Heidelberg, Germany}
  \affiliation[d]{Institute of Particle Physics and Key Laboratory of Quark and Lepton Physics (MOE), Central China Normal University, Wuhan 430079, China}
  \author[c]{Markus K. K{\"o}hler,}
  \author[e]{Aleksas Mazeliauskas,}
\affiliation[e]{Theoretical Physics Department, CERN, CH-1211 Geneva 23, Switzerland}
  \author[f]{Krzysztof Redlich,}
    \affiliation[f]{Institute of Theoretical Physics, University of Wroc\l aw, 50-204 Wroc\l aw, Poland}
  \author[b,c]{Johanna Stachel,}
 \author[g]{and Vytautas Vislavicius}
    \affiliation[g]{Niels Bohr Institute, University of Copenhagen, Copenhagen, Denmark}
\emailAdd{vytautas.vislavicius@cern.ch}

\date{\today}

\abstract{
    In relativistic nuclear collisions the production of hadrons with light (u,d,s) quarks  is quantitatively described in the framework of the Statistical Hadronization Model (SHM). Charm quarks are dominantly produced in initial hard collisions but interact strongly in the hot fireball and thermalize. Therefore charmed hadrons can be incorporated into the SHM by treating charm quarks as 'impurities' with thermal distributions, while the total charm content of the fireball is fixed by the measured open charm cross section. We call this model SHMc and demonstrate that with SHMc the measured multiplicities of single charm hadrons in lead-lead collisions at LHC energies can be well described with the same thermal parameters as for (u,d,s) hadrons. Furthermore, transverse momentum distributions are computed in a blast-wave model, which includes the resonance decay kinematics. SHMc is extended to lighter collision systems down to oxygen-oxygen and includes doubly- and  triply-charmed hadrons. We show predictions for production probabilities of such states exhibiting a characteristic and quite spectacular enhancement hierarchy.
}

\begin{document}
\maketitle
\section{Introduction}
\label{sec:Intro}

The statistical hadronization model (SHM) is the standard tool to predict and describe hadron abundances produced in relativistic nuclear collisions~\cite{Andronic:2017pug}. The main physics  assumption underlying the SHM is that, near the phase boundary between the quark-gluon plasma (QGP) at high temperature and confined hadronic matter at lower temperature, the fireball formed in such collisions is close to thermal equilibrium. In the large volume limit applicable for Pb-Pb collisions at LHC energies or Au-Au collisions at RHIC energies the produced hadrons can then be precisely described  by using a grand canonical partition function based on the hadron-resonance gas (HRG) and with residual interactions deduced using the S-matrix approach of \cite{Andronic:2018qqt}.
We note that this HRG statistical operator provides an equation of state that is very close to that emerging from lattice QCD  (lQCD) studies
in the hadronic phase ~\cite{Bazavov:2014pvz}.
Furthermore,
 the pseudo-critical temperature $T_{pc}$ at \muB = 0, which is now determined in lQCD calculations \cite{Bazavov:2018mes,Borsanyi:2020fev}
 with great precision: $T_{pc} = 156.5 \pm 1.5$ MeV~\cite{Bazavov:2018mes}, agrees within (small) uncertainties with the chemical freeze-out temperature  obtained from the SHM  analysis of light-flavour hadron production data~\cite{Andronic:2017pug,Andronic:2018qqt}.

How to extend the SHM to the charm sector, i.e., to SHMc, was outlined more than 20 years ago~\cite{BraunMunzinger:2000px} and further developed in~\cite{Andronic:2003zv,Becattini:2005hb,Andronic:2006ky,Andronic:2007zu}.
The main idea behind this development is as follows: The charm quark mass $m_c$ is much larger than $T_{pc}$ and hence thermal production of charm quarks or hadrons is strongly Boltzmann suppressed. However, with increasing center-of-mass energy the total charm production cross section which results from initial hard collisions increases strongly. If the so produced charm quarks thermalize in the hot fireball they participate in the thermal evolution as 'impurities', their total yield being determined by the charm cross section, not by the fireball temperature. Quantitatively, this is described by the charm balance equation~\cite{BraunMunzinger:2000px,Andronic:2006ky} leading to a fugacity $g_c$.  Roughly from $\sqrt{s_{NN}} > 15$ GeV on this will lead to an enhancement of hadrons with charm compared to a purely thermal description, see, e.g., Fig.~1 in~\cite{Andronic:2006ky} and the discussion below. Apart from canonical corrections~\cite{Andronic:2003zv,Andronic:2006ky} the enhancement scales $\propto (g_c)^{\alpha}$ where $\alpha$ is the number of charm quarks in a given hadron. Evidence for the thermalization of charm quarks in the fireball is discussed in~\cite{Andronic:2018qqt}. Charm quarks are deconfined inside the QGP, thermalize within the QGP and hadronize at the QCD phase boundary into open and hidden charm hadrons. This SHMc was used to predict~\cite{Andronic:2003zv,BraunMunzinger:2007zz} charmonium yields in Pb-Pb collisions at LHC energies long before the LHC turned on. It provides an excellent description of charmonium production~\cite{Andronic:2006ky,Andronic:2007bi,Andronic:2018vqh,Andronic:2019wva} without any new parameters and this success represents  compelling evidence for this new production mechanism on the hadronizing QGP phase boundary.

In the present paper we explore the predictions of the SHMc for the production of open charm mesons and baryons. Early predictions for open charm hadrons were made already in~\cite{Andronic:2003zv}, and in~\cite{Becattini:2005hb} for baryons with $\alpha > 1$, but in the absence of experimental data in the relevant low transverse momentum region these early investigations were not pursued further. The situation changed recently  when the STAR collaboration at RHIC~\cite{Adam:2019hpq} as well as the ALICE~\cite{Adam:2015sza,Acharya:2018ckj,Acharya:2020lrg,Acharya:2020uqi} and CMS~\cite{Sirunyan:2019fnc} collaborations at the LHC published first results with Au and Pb beams. It is therefore timely to provide a concise description of the SHMc in the charm sector, to compare results based on this approach to the newly available data and to extend the predictions to the multi-charm sector.  We note that the only additional information needed for SHMc predictions are the total open charm cross section and as complete as possible information on the mass spectrum of states in the charm sector. Apart from those there are no free parameters in our approach.

In Section~\ref{sec:SHM_hq} we discuss the SHMc formalism including the charm-balance equation and fugacities, the information on the total open charm cross section, and the hadron mass spectrum in the charm sector. In addition, we will lay out the framework for extending our results to lighter colliding systems of Xe-Xe, Kr-Kr, Ar-Ar and O-O, which could be studied in future runs of LHC. For the study of system size dependence of $D$ meson $R_\text{AA}$ in a dynamical heavy flavour framework see ref.~\cite{Katz:2019qwv}. For these systems and, in particular for the evaluation of production yields of multi-charm hadrons, a detailed description in terms of  canonical thermodynamics is required and outlined. This leads to thermal predictions for rapidity densities of all charmed hadrons in all colliding systems investigated here.

In section~\ref{sec:SHMc_spec} we discuss the most up-to date information of the hadron mass spectrum in the charm sector. In particular we review the theoretical and experimental motivation of additional yet-undiscovered charmed hardon states.

In section~\ref{sec:SHMc_pt} we present the description of transverse momentum spectra for charmed hadrons using a blast-wave approach. This includes a comparison of results for different freeze-out surfaces. An integral part of this approach is the incorporation of resonance decays into the calculation of spectra. In this section we also outline the 'core-corona' picture which is important to describe the high transverse momentum and centrality dependence of charm hadron production.

Results and comparisons to data are discussed in section~\ref{sec:results1}. In this section we first compare SHMc predictions to data of D-mesons and make a prediction for $\Lambda_c$ baryons. With the same approach and no new inputs aside from masses and quantum numbers of charm hadrons we show how a whole hierarchy of predictions emerges depending on whether we deal with single, double, or triple charm hadrons. Because of the above discussed enhancement of production yields for states with multiple charm these predictions will be tested in the upcoming LHC Run3 and Run4 at least for a selected number of states with $\alpha \le 2$. With a new ALICE3 experiment~\cite{Adamova:2019vkf} a large part of the whole mass spectrum of charmed mesons and baryons should be within reach. These experiments can therefore bring completely new information on the  degree of deconfinement and mechanism of hadronization of charm quarks in the hot fireball. We conclude this paper with a brief summary and outlook.

\section{Heavy quarks in the statistical hadronization model}
\label{sec:SHM_hq}

Here we recapitulate the physics ideas and formalism behind the  SHMc  with special focus on the multi-charm sector.  For more detail on the original development  see~\cite{Andronic:2003zv,Andronic:2006ky,Andronic:2017pug}. Our main emphasis will be on the description of yields and transverse momentum spectra for open charm hadrons with $\alpha \le 3$, produced in Pb-Pb collisions at LHC energy. We will also provide expressions to describe the change of yields when going to lighter collision systems including Ar-Ar and O-O and discuss briefly what can be expected. The production of charmonia or charmonium-like states has recently been investigated, see~\cite{Andronic:2017pug,Andronic:2019wva} and will not be discussed here. Our approach can also be used to make predictions for open charm hadron production at lower energies such as at the RHIC, SPS and FAIR facilities and for higher energies expected at a possible Future Circular Collider~\cite{Dainese:2016gch}.  The model can be straightforwardly extended to the beauty sector without conceptual changes or new parameters except for the total open beauty cross section and the corresponding hadronic mass spectrum. However, SHM might need to be modified for beauty hadrons, if future data reveal only partial thermalization of beauty quarks in the QCD medium.

\subsection{Multi-charm hadrons, charm balance equation and the charm fugacity factor}
\label{sec:balance}

Our starting point is the charm balance equation~\cite{BraunMunzinger:2000px}

\begin{equation}
  \begin{aligned}
    N_{\ccBar} = \frac{1}{2} & g_c V \sum_{h_{oc,1}^i} n^{{\rm th}}_i \,
    + \, g_c^2 V \sum_{h_{hc}^j} n^{{\rm th}}_j \, + \, \frac{1}{2} g_c^2 V \sum_{h_{oc,2}^k} n^{{\rm th}}_k,
  \end{aligned}
  \label{eq:balance}
\end{equation}
where $N_{\ccBar}\equiv \ud N_{\ccBar}/\ud y$ denotes the rapidity density of charm quark pairs produced in early, hard collisions and the (grand-canonical) thermal densities for open and hidden charm hadrons are given by $n_{i,j,k}^{{\rm th}}$. The index $i$ runs over all open charm states $h_{oc,1}^i = D, D_s, \Lambda_c, \Xi_c, \cdots, \bar{\Omega}_c$ with one valence charm or anti-charm quark, the index $j$ over all hidden charm states $h_{hc}^j = J/\psi, \chi_c, \psi',\cdots$, and the index $k$ over open charm states $h_{oc,2}^k = \Xi_{cc} \cdots, \bar{\Omega}_{cc}$ with two charm or anti-charm quarks. We leave out here states with 3 charm or anti-charm quarks as their contribution to the sum is negligible for realistic masses and values of $g_c$ and they have yet to be discovered. These thermal densities are computed using the latest version of the SHMc~\cite{Andronic:2017pug,Andronic:2019wva}
with the chemical freeze-out temperature $T_{cf}= 156.5$ MeV and the fireball volume per unit rapidity at mid-rapidity $V = 4997\pm 455\,\text{fm}^3$ as appropriate for the most central 10\% Pb-Pb collisions at LHC energy $\sqrt{s_{NN}}= 5.02$ TeV. In the appendix we also give results for the 30-50\% centrality interval and at mid-rapidity. Scaling with the measured charged particle pseudo-rapidity density the corresponding volume in this centrality bin is  $V = 1238\pm 113\,\text{fm}^3$. For the results shown below, the uncertainties in volume were not propagated, because they are sub-leading compared to the uncertainty in $g_c$ discussed below.

The total number of charm quark pairs $N_{\ccBar}$ produced in a Pb-Pb collision is a quantity that should be determined by measurement of all hadrons with open or hidden charm. Following this prescription, the only (additional) input parameter of the SHMc, $N_{\ccBar}$, is determined by experiment. In particular, we note that $N_{\ccBar}$ already includes all nuclear effects in charm production as compared to pp collisions, takes into account potential additions to the charm yield from thermal production in the QGP as well as potential losses due to charm quark annihilation.
In practice, using this prescription is, however, difficult since the measurement of all open and hidden charm hadrons needs to be performed without cuts in transverse momentum. Achieving a precision measurement of $N_{\ccBar}$ is one of the priorities for the upgraded ALICE experiment in LHC Run3 and Run4.

In the absence of a measured charm production cross section in Pb-Pb collisions we obtain $N_{\ccBar}$ at mid-rapidity from the measured  charm cross section $\der \sigma_{c\bar{c}}/\der y$ in pp collisions by multiplication with the appropriate nuclear thickness function for Pb-Pb collisions and taking into account nuclear modifications. The procedure is described in detail below.

The pp data were measured at $\sqrt{s}= 5.02$ and 7 TeV at mid-rapidity~\cite{Adam:2016ich,Acharya:2017jgo,Acharya:2019mgn,Acharya:2019mno}. To apply to Pb-Pb collisions, the cross sections are multiplied with the nuclear thickness function and folded with a factor accounting for nuclear modification effects such as shadowing, energy loss or saturation effects. The estimate of this factor is based on the analysis of prompt $D^0$ and \jpsi production in p-Pb collisions at 5.02 and 8.16 TeV. We used the data from the LHCb collaboration~\cite{Aaij:2016jht,Aaij:2017cqq,Aaij:2017gcy} at forward rapidity, and of \jpsi production at mid-rapidity measured by the ALICE collaboration in pp and p-Pb collisions at 5.02 TeV~\cite{Acharya:2019mgn,Acharya:2019mno}. The $\sqrt{s}= 8.16$  and 7.0 TeV data are interpolated to 5.02 TeV using the measured data at other center-of-mass energies and the functional form obtained from perturbative QCD (FONLL) ~\cite{Cacciari:2015fta}. For mid-rapidity, we obtain a reduction factor of $0.65 \pm 0.12$, resulting in a value of $\der \sigma_{\ccBar}/ \der y = 0.532 \pm 0.096$~mb. The corresponding factor for $y$ = 2.0-4.5 is $0.70 \pm 0.08$ leading to a differential charm production cross section of $\der \sigma_{\ccBar}/ \der y = 0.334 \pm 0.053$~mb. To obtain the charm quark rapidity density for Pb-Pb collisions of a given centrality, the pp cross section is then multiplied with the mean nuclear thickness function $\left<T_\text{AA}\right>$ as described in~\cite{Abelev:2013qoq}.  We neglect in the procedure based on results from pp and p-Pb collisions potential contributions to the differential charm cross section in Pb-Pb collisions from thermal charm production as well as reductions from charm quark annihilation. For LHC  both contributions were estimated to be very small and negligible for lower energies~\cite{BraunMunzinger:2000dv,Andronic:2006ky}.

We note here that the charm balance equation should contain canonical corrections for more peripheral collisions or for lighter collision systems, i.e., whenever the number of charm pairs is not large compared to 1~\cite{Gorenstein:2000ck,BraunMunzinger:2003zd}.
The charm balance Eq.~\ref{eq:balance} needs then to be modified accordingly. To that end we define
\begin{equation}
\begin{aligned}
 N_{oc,1} = \frac{1}{2} g_c V \sum_{h_{oc,1}^i} n^{{\rm th}}_i,\\
 N_{oc,2} = \frac{1}{2} g_c^2 V \sum_{h_{oc,2}^k} n^{{\rm th}}_k,\\
 N_{hc} =  g_c^2 V \sum_{h_{hc}^j} n^{{\rm th}}_j,
 \label{eq:charm_numbers}
 \end{aligned}
\end{equation}
where $N_{oc,1}$ is the rapidity density of charm quarks bound in hadrons $h_{oc,1}^i$ with one valence charm quark, $N_{oc,2}$ is the rapidity density of charm quarks bound in hadrons $h_{oc,2}^k$ with two valence charm quarks, and $N_{hc}$ is the rapidity density of charm-(anti-charm) quark pairs bound in hidden charm hadrons $h_{hc}^j$. This defines the total rapidity density of charm quarks, neglecting triply charmed states, as $N_c^\text{tot} = N_{oc,1} + N_{oc,2} + N_{hc}$. Note that the value of $N_c^\text{tot}$ itself depends on the charm fugacity $g_c$. Then the modified charm balance equation using the canonical corrections reads:

\begin{equation}
N_{c\bar{c}}= \sum_{\alpha = 1,2} N_{oc,\alpha} \frac{I_\alpha(N_c^\text{tot})} {I_0(N_c^\text{tot})} \, + \, N_{hc}.
\label{eq:canonical}
\end{equation}
Here, the $I_\alpha$ are modified Bessel functions. For hadrons with 2 or 3 charm quarks there are generally additional terms which are, however, very small because of the small charm densities, and are neglected here (see, e.g. sect. 3.2 in~\cite{BraunMunzinger:2003zd}).

Solving Eq.~\ref{eq:canonical} for $g_c$ then determines the charm fugacity factor at 5.02 TeV. For central (0-10\%) Pb-Pb collisions and the above discussed differential charm cross section at mid-rapidity (implying $\ud N_{c\bar{c}}/\ud y$=12.95$\pm$2.27) this leads to $g_c = 29.6 \pm 5.2$, with the uncertainty determined by the uncertainty in the open charm cross section for Pb-Pb collisions. The rapidity density of open charm hadrons of type $ h_{oc,\alpha}^i $ with $\alpha=1,2$ charm quarks can then be obtained from the computed thermal densities $n_{i}^{\rm th}$ as :
\begin{equation}
   \frac{\ud N(h_{oc,\alpha}^i)}{\ud y} =g_c^\alpha \, V \, n^{{\rm th}}_i \frac{I_{\alpha}(N_c^\text{tot})}{I_0(N_c^\text{tot})}.
\label{eq:yieldsoc}
\end{equation}
The large value of $g_c = 29.6 \pm 5.2$  for central Pb-Pb collisions
for charm production at mid-rapidity (see Fig.~\ref{fig:gc-scaling} in the following section) implies very large enhancements for charmed hadrons compared to what is obtained in the purely thermal case. In the absence of canonical corrections the enhancement factor is (nearly) 900 for doubly charmed, and $ 2.6 \cdot 10^4$ for triply charmed hadrons. For central Pb-Pb collisions at 5.02 TeV the canonical correction factors are in fact close to 1: 0.98, 0.92, and 0.84 for $\alpha = 1, 2, 3$ charm quarks respectively, for the central value of the differential charm cross section at mid-rapidity, see Fig.~\ref{fig:canonical} below. If these enhancement factors are realized in nature then even very massive triply charmed hadrons may come into reach experimentally.

For hidden charm states Eq.~\ref{eq:yieldsoc} reduces to
\begin{equation}
    \frac{\ud N(h_{hc}^j)}{\ud y} = g_c^2 \, V \, n^{{\rm th}}_j.
\label{eq:yieldshc}
\end{equation}

The enhancement factors expressed in Eqs.~\ref{eq:yieldsoc} and \ref{eq:yieldshc} come about because of the assumption that all charm quark reach thermal equilibrium at least for temperatures close to $T_{cf}$. In that case the heavy quarks are completely uncorrelated and the resulting statistical weight is just $g_c^\alpha$. We note that this implies deconfinement of the heavy quarks over the volume $V$, as discussed below.

We also stress that all hadron rapidity densities discussed above are computed as rapidity densities for a volume and hence rapidity window of width of $\Delta y =1$. The rationale behind this is that one cannot combine charm quarks into hadrons over large rapidity distances as they are causally disconnected because hadrons have a finite formation time $\tau_f \approx 1$ fm and large rapidity correlations can only be established at very early times $\tau \ll 1$ fm~\cite{Acharya:2019izy,Dumitru:2008wn}. The value of $\Delta y$ is somewhat arbitrary and a range of $\Delta y = 1 - 3$ was explored in the past and for colliders a weak dependence was found \cite{Andronic:2003zv}. We finally note the asymptotic form of the modified Bessel functions $I_\alpha(x)$. For small argument $x$ and order $\alpha$ this reads:
\begin{equation}
  I_{\alpha}(x)  \approx \frac{1}{\Gamma(\alpha + 1)} (x/2)^{\alpha}
\label{eq:bessel}
\end{equation}
where $\Gamma$ is the Euler Gamma function. For large $x$ the modified Bessel functions approach
\begin{equation}
 I_\alpha(x) \approx \frac{e^x}{\sqrt{2\pi x}}.
 \label{eq:bessel1}
\end{equation}
This implies that the canonical suppression disappears for large arguments $x$, i.e., the system has reached the grand-canonical limit. For small $x$, $I_0 \approx 1$ and the canonical suppression factor approaches $\frac{1}{\Gamma(\alpha + 1)} (x/2)^{\alpha}$.

\subsection{Dependence on mass number of the colliding nuclei}
\label{sec:A-dependence}

In the following we provide information on how to also compute the yields for (multi\nobreakdash-)charm hadrons produced in lighter collision systems such as Xe-Xe,  Kr-Kr, Ar-Ar and O-O. Of course, these calculations  are valid as long as the charm quarks produced in initial hard collisions reach or closely approach kinetic equilibrium in the hot fireball formed in the collision. This has to be carefully checked when one plans to study the production of charm hadrons in such small systems. In addition, we have not included in these exploratory calculations any contributions due to corona effects. Their importance will increase as the colliding systems become smaller. For the system O-O where the nuclear densities never reach a central plateau we expect very substantial corrections which need to be studied carefully if one wants to look for QGP effects in such very light systems. For more discussion on the corona effect see section~\ref{sec:SHMc_pt} below.

\begin{figure}
\centering
\includegraphics[scale=0.35]{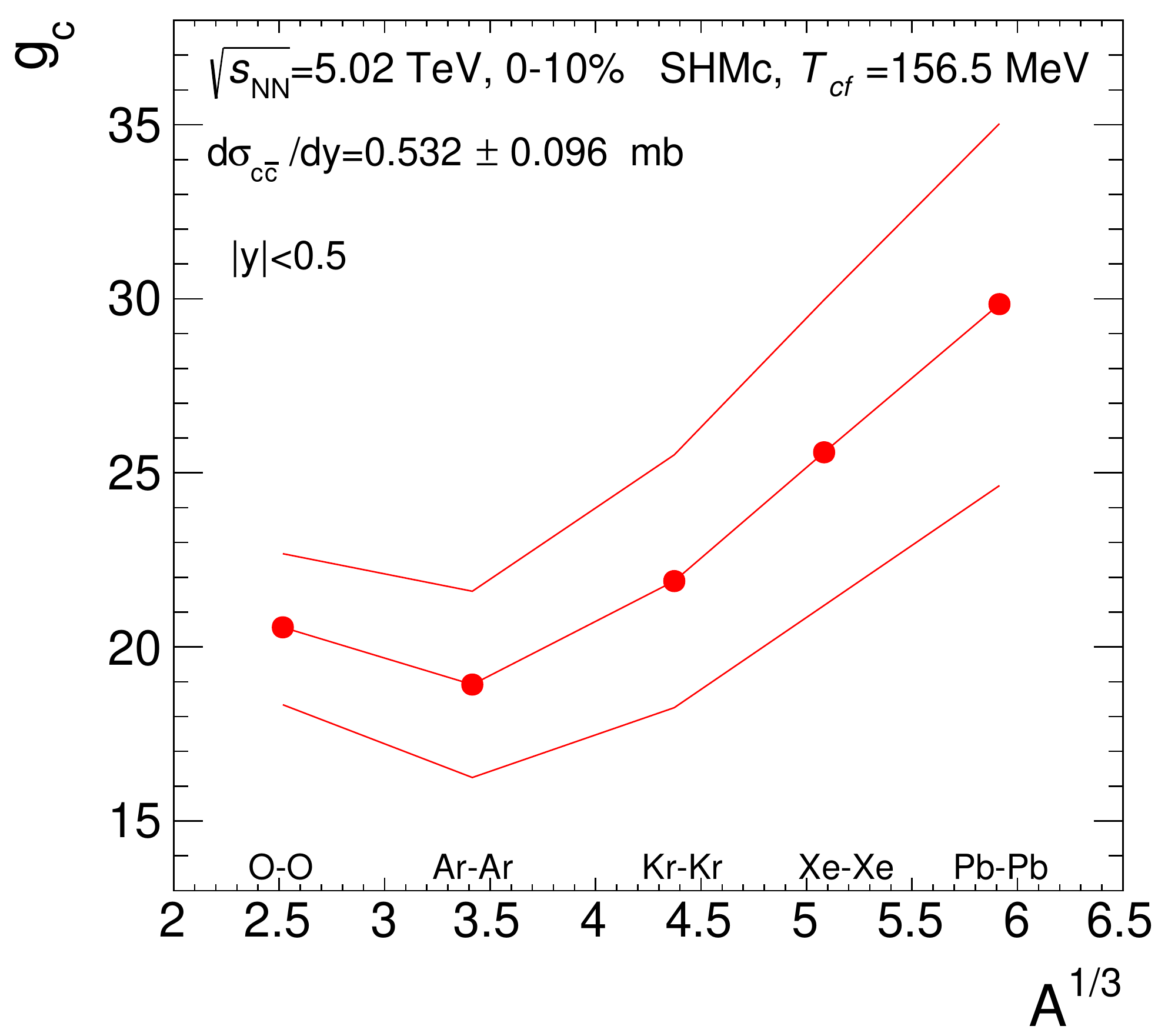}
\includegraphics[scale=0.35]{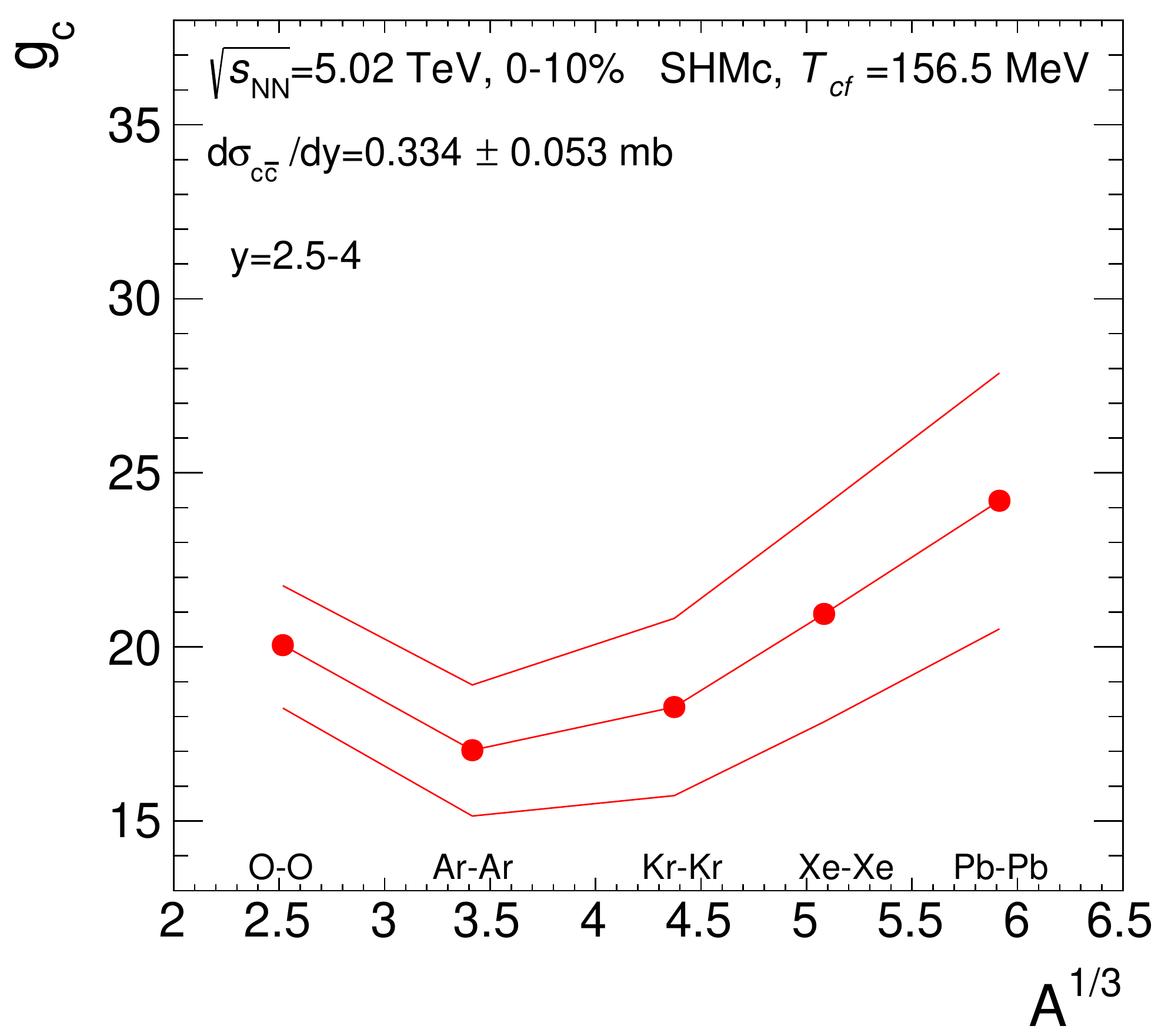}
\vskip -0.4 cm
\caption{The system-size (expressed as $\mathrm{A}^{1/3}$) dependence of the charm fugacity factor $g_c$ for the five different collision systems Pb-Pb, Xe-Xe, Kr-Kr, Ar-Ar, and O-O for rapidity $|y| < 0.5$ (left plot) and rapidity 2.5-4 (right plot). The band reflects the uncertainties of $\ud\sigma_{c \bar c}/\ud y$ indicated in the plots.
For details see text.}
\label{fig:gc-scaling}
\end{figure}

To understand the charm hadron yield dependence on mass number A of the colliding nuclei we first determine the A dependence of $g_c$. From the charm balance Eqs.~\ref{eq:balance} and~\ref{eq:canonical} we note that $N_{c\bar{c}} \propto {\rm A^{4/3}}$ since charm is produced in hard collisions and we are interested in central nuclear collisions~\cite{dEnterria:2003xac}. Noting further that the volume $V \propto$ A we immediately obtain that $g_c \propto {\rm A^{1/3}}$ in the grand-canonical limit. In the canonical limit, i.e., for small charm densities, one obtains $g_c \propto {\rm A^{-1/3}}$ using the properties of the modified Bessel functions near the origin (see Eqs.~\ref{eq:bessel} and \ref{eq:bessel1}). However, at LHC energies charm densities are not so small and the grand-canonical approximation is a good approximation for the heavier systems Xe-Xe and Kr-Kr and leads to a 20\% correction for Ar-Ar. The correction becomes large for the O-O system. In Fig.~\ref{fig:gc-scaling} we show the result of the A dependence of $g_c$ as obtained by numerical solution of Eq.~\ref{eq:canonical}.

\begin{figure}
\centering
\includegraphics[scale=0.35]{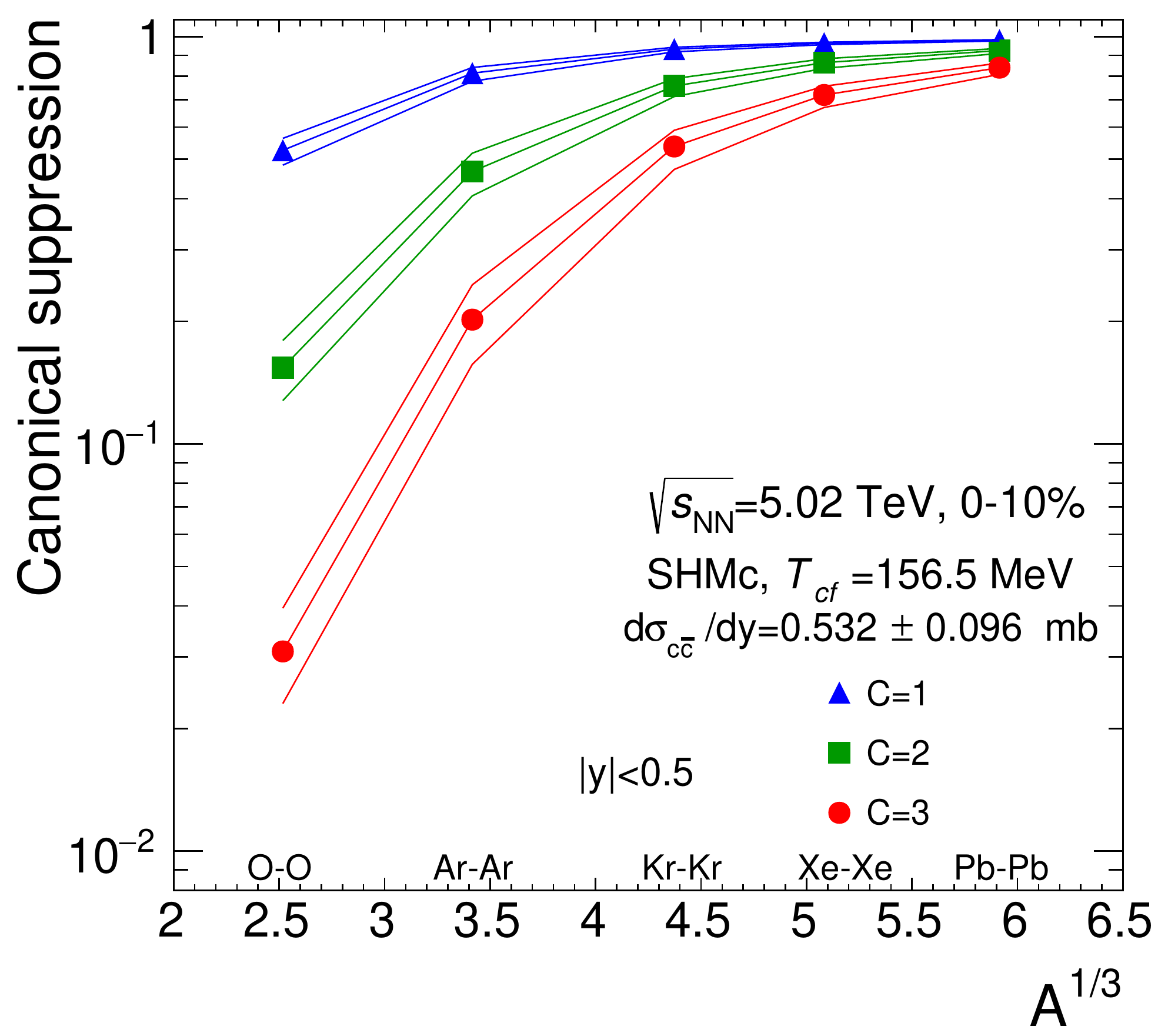}
\includegraphics[scale=0.35]{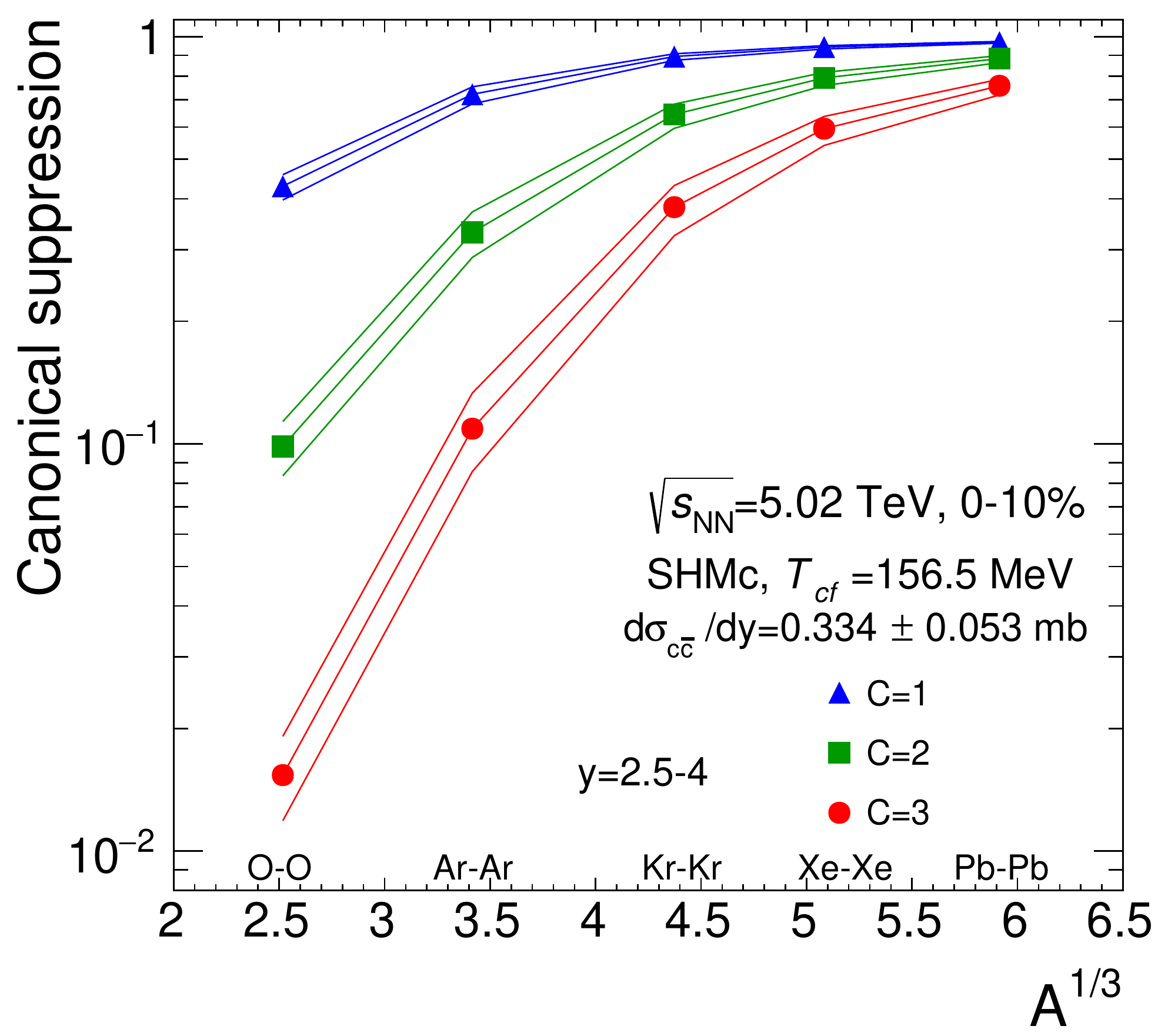}
\vskip -0.4 cm
\caption{Canonical correction factors for the five different collision systems Pb-Pb, Xe-Xe, Kr-Kr, Ar-Ar, and O-O at mid-rapidity $|y| < 0.5$ (left panel) and forward rapidity 2.5-4 (right panel) for open flavor hadrons with  charm quantum number C. The bands reflect the uncertainties of $\der \sigma_{c \bar c}/\der y$ as indicated in the figure. For details see text.}
\label{fig:canonical}
\end{figure}

The rather strong deviation from the ${\rm A^{1/3}}$ dependence observed for the O-O system is caused by the changes in the canonical correction factor due to the transition from grand-canonical to canonical thermodynamics where the A dependence of $g_c$ is expected to approach the ${\rm A^{-1/3}}$ scaling as discussed above. For the rapidity range 2.5-4 the non-monotonic feature of the curves is more pronounced,  as the system is deeper into the canonical regime, see Fig.~\ref{fig:canonical}.

In Fig.~\ref{fig:canonical} we present the dependence on mass number A of the canonical correction factors $f_{can}$ for the production of charm hadron $h^i$ in A-A collisions. They are defined as:
\begin{equation}
f_{can}(\alpha,{\rm A}) =  \frac{I_{\alpha}(N_c^\text{tot}({\rm A}))}{I_0(N_c^\text{tot}({\rm A}))}.
\label{eq:f_can}
\end{equation}
The curves on the left and right side are again obtained at rapidity $|y| < 0.5$ and rapidity 2.5-4, respectively. They are evaluated for charm hadrons with the expression given in equation~\ref{eq:canonical}.  The A dependence of $g_c$ needs to be obtained numerically and is displayed in Fig.~\ref{fig:gc-scaling} above.

With the A-dependence of $g_c$ and of the canonical corrections factors at hand we can now compute the yield of any charmed hadron in the SHMc as function of mass number A. In section~\ref{sec:results1} below we will present our results on yields and transverse momentum distributions.

To get a more intuitive understanding of these results we assume, in the following, that the A dependence of $g_c$ can be  described by the above grand-canonical relation $g_c \propto {\rm A^{1/3}}$. As can be seen from Fig.~\ref{fig:gc-scaling}, this is well fulfilled, at the better than 10\% (1\%) level, for A $\ge$ 40 (80).  Keeping these small deviations in mind, we can provide a good estimate of the A dependence of charm hadron yields provided we stay with A $\ge$ 40 , i.e., Ar-Ar collisions, by making use of Eq.~\ref{eq:yieldsoc} and the above defined canonical suppression factors $f_{can}$. This  leads to the scaling relation
\begin{equation}
    \frac{\der N^{\rm AA}}{\der y}(h^i)=\frac{\der N^{\rm PbPb}}{\der y}(h^i) \left(\frac{{\rm A}}{208}\right)^{(\alpha+3)/3} \frac{f_{can}(\alpha,{\rm A})}{f_{can}(\alpha,{\rm Pb})}
    \label{eq:scaling}
\end{equation}
for the production of hadron $h^i$ with $\alpha$ charm quarks in collision systems of A-A. Using this relation and the yields for charm hadrons produced in Pb-Pb collisions as displayed in Table~\ref{tab:yields_tot}, see section~\ref{sec:results1} below, the yields can be computed for charm hadrons yields in lighter systems from Ar-Ar to Xe-Xe. For very light systems such as O-O the full approach as discussed above should always be used.

\begin{figure}
\centering
\includegraphics[scale=0.35]{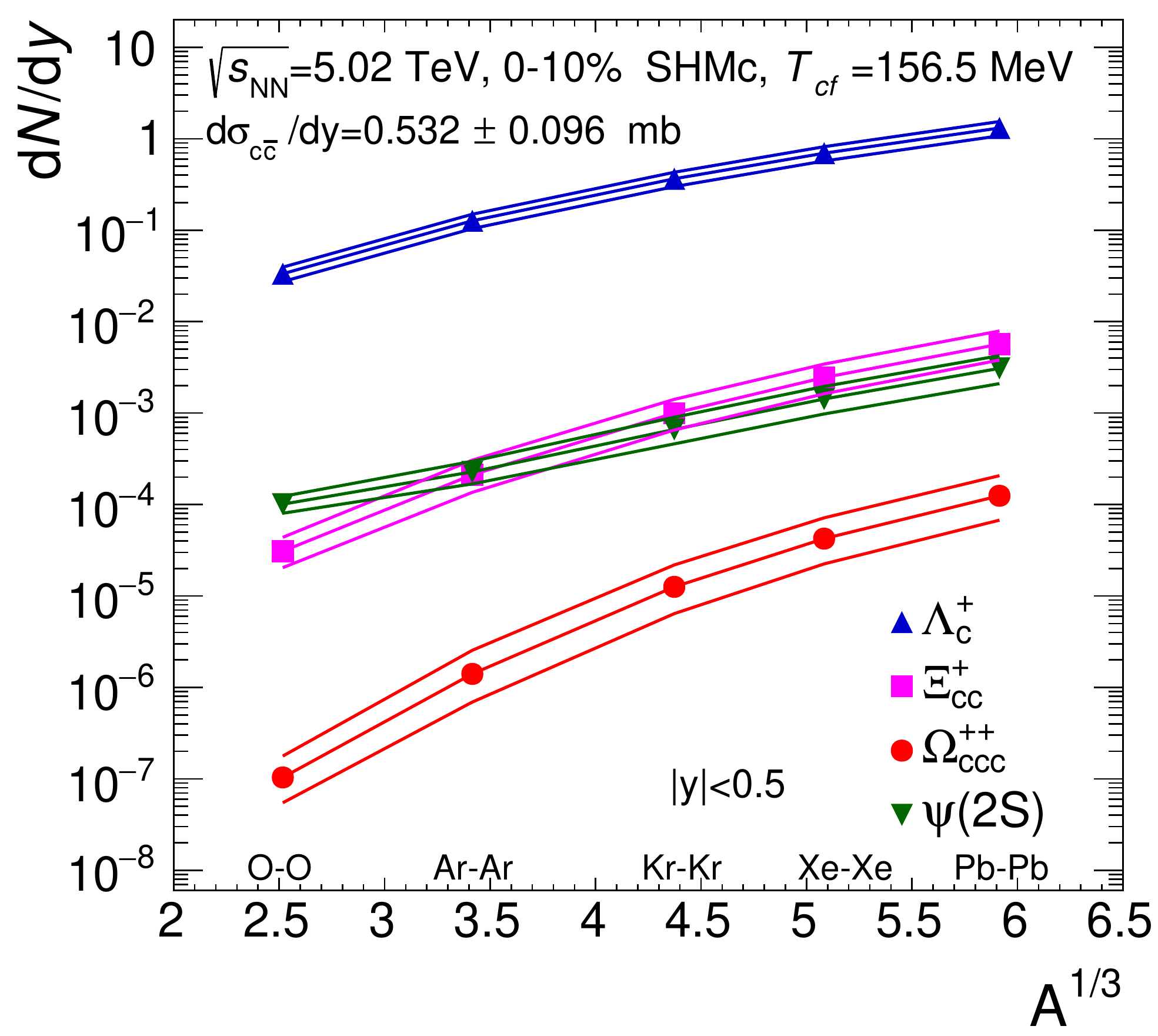}
\includegraphics[scale=0.35]{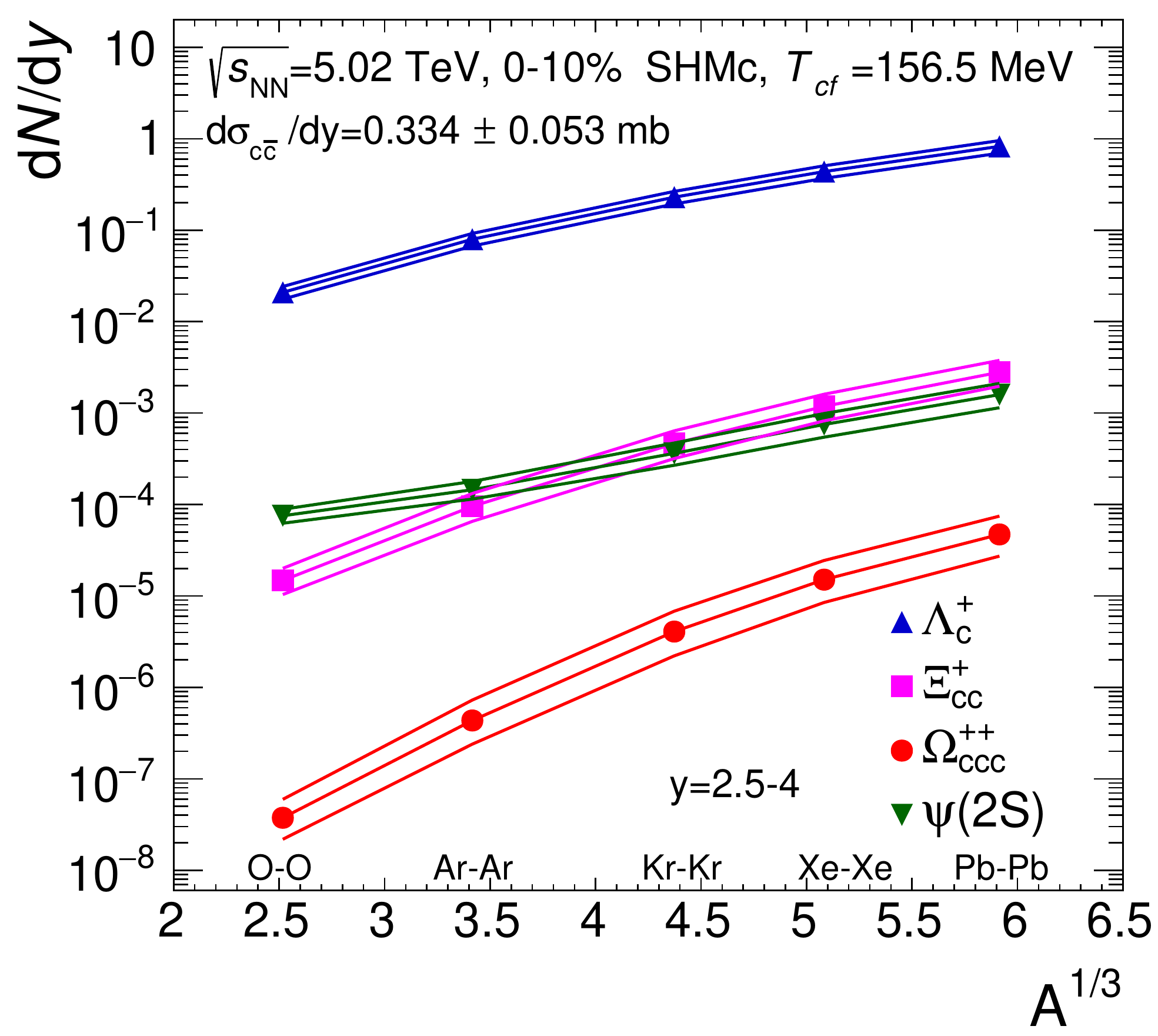}
\vskip -0.4 cm
\caption{System size dependence of selected hadron species for mid-rapidity $|y| < 0.5$ (left panel) and forward rapidity 2.5-4 (right panel).}
\label{fig:yields_a}
\end{figure}

In Fig.~\ref{fig:yields_a} the system size dependence of selected hadron yields is displayed for mid-rapidity (left panel) and forward rapidity (right panel). The band for each hadron species correspond to different charm production cross sections as indicated in the figure. Note the change in A dependence for open and hidden charm states as a consequence of the absence of the canonical suppression for the latter (compare Eq.~\ref{eq:yieldshc} and \ref{eq:yieldsoc} above).

\subsection{The canonical volume}
\label{sec:can_vol}

The volume $V$ appearing in Eq.~\ref{eq:balance} is usually set equal to the fireball volume at chemical freeze-out $V$ determined by the requirement that the measured rapidity density of charged particles divided by $V$ equals the thermal density of charged particles after strong decays at chemical freeze-out~\cite{Andronic:2017pug}. Employing a connection between momentum rapidity and space-time rapidity, this volume, corresponding to one unit of rapidity, is a fraction of the entire fireball. To consider such a sub-volume is meaningful since, at high collision energies, equilibration is achieved only locally and not globally. This leads to the picture at freeze-out of a string of fireballs lined up in rapidity and filling the entire gap between the rapidities of the two beams (or between beam and target in fixed target mode). The thermal parameters of these fireballs could differ, albeit at LHC we expect a slow variation with rapidity. Only at low collisions energies (AGS energy and below) one should think of one global thermalized system. We note in this context that in~\cite{Becattini:2005hb} it was assumed that the fireball volume comprises all rapidities up to but excluding beam and target rapidities, hence is significantly larger than what is discussed here.

When computing the canonical suppression factor $f_{can}$ defined in Eq.~\ref{eq:f_can}, a new scale enters the problem. To obtain the argument of the Bessel functions, the differential cross section or multiplicity needs to be multiplied with the width of a rapidity interval $\Delta y$ which then can be associated with a canonical volume $V_{can}$ over which the relevant quantum number is conserved. For the conservation of baryon number we have recently learned, in the context of net-proton fluctuations, that this volume $V_{can}$ may be significantly larger, not smaller than $V$~\cite{Braun-Munzinger:2019yxj,Acharya:2019izy}. Very recent results concerning canonical strangeness suppression~\cite{Cleymans:2020fsc} at the LHC point also in that direction.
Since charm quarks are all produced in the very early phase of the collision we could expect that the canonical volume for charm $V_{can}$ is similarly large, implying a reduced role of canonical suppression and yields larger than computed with $V = V_{can}$. This would affect  in particular predicted yields for multi\nobreakdash-charm hadrons from lighter collision systems such as Ar-Ar or O-O. In the numbers given below for (multiple) charm production yields canonical suppression is included. To stay on the conservative side and in the absence of measurements of $V_{can}$ for charm we have, in the following employed only one volume setting $V_{can} = V$, implying that the canonical corrections for the smallest collision systems could be less severe when more information on $V_{can}$ becomes available.

\subsection{Charm hadron production and deconfinement of charm quarks}
\label{sec:deconfinement}
Early on it was realized~\cite{Andronic:2003zv,Andronic:2007bi,BraunMunzinger:2009ih} that a successful description of the measured yields of charmonia  in the SHMc would imply deconfinement for charm quarks. The measurements at RHIC and, in particular, LHC energy lend support to this interpretation~\cite{Andronic:2017pug}. Here we briefly discuss what could be learned on deconfinement from analysis of multi-charm meson and, in particular, baryon production data.

In the SHMc the production of hadrons with $\alpha$ charm quarks is enhanced by a factor $(g_c)^{\alpha}$ compared to what is expected in a purely thermal approach, see Eq.~\ref{eq:yieldsoc}. Since $g_c \approx 30$ for central Pb-Pb collisions, the expected enhancements for multi-charm hadron production are very substantial and produce a distinctive hierarchy in their yield pattern, as shown below. That pattern results only if the charm quarks making up the final hadron are uncorrelated prior to hadronization as is expected for fully deconfined ('no strings attached') charm quarks. We note that even the residual correlation imposed by overall baryon number and charm conservation will be very small if the measurement window is of order one unit in rapidity~\cite{Acharya:2019izy}.

Production of multi-charm hadrons in the (confined) hadronic phase would also be very small as it would necessarily have to involve exotic multi-particle collisions. To illustrate this point, the following estimates are based on energy conservation and on masses of 4.8 GeV for $\Omega_{ccc}$~\cite{Zhao:2020jqu} and 3.62 GeV for $\Xi_{cc}$~\cite{Zyla:2020zbs}. For the most exotic case of $\Omega_{ccc}$ production a possible production path is via collisions such as $3D + m\pi \rightarrow \bar{p} + \Omega_{ccc}$ with $m$ = 3.  For the $\Xi_{cc}$ baryon the analogous rate equation reads  $2D + m\pi \rightarrow \bar{p} + \Xi_{cc}$ with $m$ = 7. But many other processes such as $\Lambda_c + D\rightarrow \Xi_{cc} + \pi$ or $\Lambda_c+ 2D\rightarrow  \Omega_{ccc} + \pi$ are imaginable.  While the rates for all these processes will be enhanced compared to purely thermal estimates by a fugacity factors $(g_c)^{\alpha}$, they will, nevertheless, be very small because of the low $D$ meson and $\Lambda_c$ density of $1.2 \cdot 10^{-3}\,\text{fm}^{-3}$ (for $D^0$, the highest for $D$ mesons) and $ 2.6 \cdot 10^{-4}\,\text{fm}^{-3}$
for $g_c = 29.6$ at chemical freeze-out entering at the same power of $\alpha$. These rates will fall very rapidly with temperature during the hadronic expansion~\cite{BraunMunzinger:2003zz}. Also the phase after chemical freeze-out is by construction not in equilibrium. How to constrain the rate for such multi-particle collisions is totally unclear due to the unknown amplitudes for these different possible many-body collision processes. Similar arguments apply for charmonia, where the dominant channel would be $D + \bar{D} \rightarrow J/\psi + \pi$. Here, even the extension to $\psi'$ involves at least one more unknown parameter.
This is to be contrasted with the SHMc approach where there is no free parameters. The experimental observation of a significant number of hadrons with multiple charm  in relativistic nuclear collisions hence provides a unique opportunity to test the 'deconfinement' prediction and get quantitative information on the degree of deconfinement achieved in the hot fireball.

The full predictions of the model,  including the contribution from the low density corona, are presented for a selection of species in Table~\ref{tab:yields_tot} for Pb-Pb collisions at 5.02 TeV, for the 0-10\%  and 30-50\% centralities (mid-rapidity values). For these hadrons, the production cross sections in pp collisions have recently been measured by ALICE at mid-rapidity \cite{Acharya:2021cqv,Acharya:2019mgn,Acharya:2020lrg,Acharya:2019lkw} and those are employed for the calculation of the corona component (we have employed the ratio $\psi(2S)/(J/\psi)$=0.15 \cite{Andronic:2017pug}).
The model predictions for the core part for all systems for the two rapidity ranges are available in numerical form as auxiliary file with the arXiv version of the publication.

\section{Charm hadron spectrum and SHMc}
\label{sec:SHMc_spec}

The spectrum of open charm hadrons incorporated in the SHMc includes all mesons and baryons established experimentally as given by the PDG \cite{Zyla:2020zbs}. This includes 27 D mesons and their anti-particles with angular momenta from 0 to 3 and masses up to 3 GeV. There are 36 established singly-charmed baryons and as many anti-baryons in the mass range up to 3.12 GeV. The known angular momenta are low, mostly 1/2 and 3/2 with one established 5/2 state. The thermal population of the charmed hadrons is strong enough so that the density of the ground state $D^0$ is quadrupled due to feeding from strong decays, the $\Lambda_c$ density is increased by a factor 5 due to feeding.
There has been discussion recently that the number of charmed baryons, in particular, could be significantly larger. Fourth order susceptibilities were constructed and evaluated in lQCD calculations \cite{Bazavov:2014yba} and compared to results from HRG calculations of the same quantities in the temperature range up to the pseudo-critical temperature. The ratios were chosen such that they are particularly sensitive to contributions from the charmed baryon sector in the HRG. It was found that the lQCD results are significantly (at least 40\%) above the HRG calculation based on the states established by the PDG in 2012, while adding to the HRG charmed baryon states obtained from a lQCD calculation \cite{Padmanath:2013bla}, resulted in good agreement up to the pseudo-critical temperature. The authors of \cite{Bazavov:2014yba} view this as evidence for so far unobserved charmed hadrons contributing to the thermodynamics in the cross over region. Indeed, while the spectrum of \cite{Padmanath:2013bla} is consistent with the number of known states in the mass range above the respective ground state, about 200 additional baryons with total angular momenta up to 7/2 are predicted. Most of these states are significantly higher in mass. For the positive parity states there is a mass gap of about 500-600 MeV, the gap is only of the order of 400 MeV for the negative parity states (that are generally about 300 MeV higher in mass). The situation is only different for the negative parity $\Xi_c$ states, where the new states start right at the mass of the highest experimentally established state at 3123 MeV. Accordingly, at a freeze-out temperature $T_{cf}= 156.5$ MeV the thermal weights are significantly lower. Still, due to their large number and in part also higher degeneracy factors the feeding of ground state charmed baryons could be significantly affected. In this context it is interesting to note that a wealth of new XYZ states were found at the LHC while only 1 additional $\Lambda_c$, 2 $\Xi_c$ and 5 $\Omega_c$ states were newly discovered (compare e.g. the PDG2012 and PDG2020 compilations).

Triggered by the surprizingly large fragmentation of charm into $\Lambda_c$ measured in pp collisions at 7 and 5.02 TeV by the ALICE collaboration \cite{Acharya:2017kfy,Acharya:2020uqi,Acharya:2020lrg}, He and Rapp \cite{He:2019tik} incorporated into a SHM calculation a hadron spectrum resulting from a relativistic quark model calculation \cite{Ebert:2011kk} exhibiting a very large number of additional charmed baryons with angular momenta up to 11/2 and both parities. The additional charmed baryons from the RQM calculation have by and large smaller masses than resulting from lQCD \cite{Padmanath:2013bla}, falling in part even into the mass range of the known states. Using this charmed baryon spectrum and a temperature of 170 MeV, the authors of \cite{He:2019tik} find a doubling of the $\Lambda_c$ ground state population as compared to the PDG spectrum and predict a yield in line with the ALICE experimental data.

It should be noted that this poses a conceptual problem because it implies that charmed baryons exist at a temperature significantly above the pseudo-critical temperature for the chiral phase transition, while this is explicitly not supported by lQCD calculations. In \cite{Bazavov:2014yba} it is argued that cumulants on net charm fluctuations indicate that above $T_{pc}$ the charm degrees of freedom are no longer described by an uncorrelated gas of charmed hadrons but that rather the emergence of deconfined charm states sets in just near the chiral cross over transition. On the other hand, Petreczky \cite{Petreczky:2020olb} notes that while the ratio of fourth order baryon-charm susceptibilities around and above the pseudo-critical temperature of the chiral transition is much above the values for the HRG but still below the free quark gas value, that fact could be understood if charm hadron like excitations would still exist above $T_{pc}$ possibly up to 200 MeV. This is not the baseline of the predictions of this publication where deconfinement of all flavors at $T_{pc}$ is assumed. The predictions presented below will provide a stringent test of charm deconfinement and settle this discussion once a large enough dynamic range in mass and charm quantum number is covered by experimental data. Finally we quote recent lQCD results \cite{Lorenz:2020uik} where comparisons of Euclidean correlators to perturbative spectral functions were found to be indicative of charmonium melting in lQCD very close to $T_{pc}$.

While the questions raised here are debated in the community, we want to give an indication in this publication how the SHMc predictions given below would be affected by a large number of yet undiscovered charmed baryons behaving like simple resonances. To this extent we have performed also calculations where the statistical weight of all excited charmed baryons was tripled and the corresponding change in the predictions by the SHMc is given in section \ref{sec:results1} where hadron yields are presented. Finally it should be noted that, even if the above plethora of charmed baryons exists, a treatment as simple resonances in the SHMc could be too naive and a situation could arise similar to the light quark sector. In a recent study~\cite{Andronic:2020iyg}, the SHM was augmented by 180 nonstrange and 300 strange baryons predicted by lQCD. When they were treated a simple additional resonances, their presence showed a significant impact on particularly the proton yield, strongly deteriorating agreement with experimental data. Proper treatment of the pion-nucleon interaction by the S-matrix approach and using all measured phase shifts \cite{Andronic:2018qqt} completely cancelled out the effect of these additional states. This strong effect of the S-matrix approach could be traced \cite{Lo:2017lym} to non-resonant and repulsive components in the pion-nucleon interaction for some partial waves. Whether such a situation could arise in the charm baryon sector depends, among other things, on the widths of the additional states, and is currently completely unexplored. We have assumed that all additional resonances are narrow Breit-Wigner-type  resonances.

\section{Transverse momentum spectra of charm hadrons}
\label{sec:SHMc_pt}
In the  SHM fitted to integrated particle yields no assumption is made  about the form of the momentum spectra of produced particles. Therefore the transverse momentum dependence  must be supplied by additional modelling of the particle freeze-out.

In hydrodynamical modelling of heavy ion collisions the soft momentum part of particle spectra is obtained by the Cooper-Frye~\cite{Cooper:1974mv} integral over the freeze-out surface and subsequently passing to the hadronic afterburner to perform resonance decays and possible hadronic rescattering.
The blast-wave model~\cite{Schnedermann:1993ws,Florkowski:2010zz} is motivated by the same physics picture, but realized in simpler but approximate way to generate the $\pT $ spectra. The thermal particle spectra are obtained from a simple freeze-out surface with a given freeze-out temperature and with  parametrized radial velocity profile. This thermal blast-wave model has been used extensively in the past to fit and characterize the experimentally measured identified particle spectra~\cite{Abelev:2013vea,Acharya:2019yoi,Acharya:2020zji,Acharya:2018orn}.

For boost-invariant and azimuthally symmetric freeze-out surfaces $d\sigma_\mu$, the Cooper-Frye integral can be reduced to a one-dimensional integral along the freeze-out contour in the  $\tau$-$r$ plane~\cite{Schnedermann:1993ws,Florkowski:2010zz}:
\begin{align}\label{eq:Cooper-Frye}
&  \frac{\ud^2 N}{2\pi \pT d\pT dy} =\frac{2J+1}{(2\pi)^3}\int \ud \sigma_\mu p^\mu f(p)\nonumber\\
  &= \frac{2J+1}{(2\pi)^3} \int_0^{r_\text{max}}\!\! \ud r \; \tau(r) r  \left[ K^\text{eq}_1(\pT ,u^r) -  \frac{\partial \tau}{\partial r}  K^\text{eq}_2(\pT ,u^r) \right],
\end{align}
where $2J+1$ accounts for spin-degeneracy.
Here we consider a freeze-out surface defined by a single-valued function $\tau(r)$ in the range $0<r<r_\text{max}$.

The freeze-out kernels $K^\text{eq}_{1,2}(\pT ,u^r)$ can be calculated analytically for the Boltzmann distribution $f(p) = \exp(-\sqrt{m^2+p^2}/T)$ of initial particles on the freeze-out surface and takes the well-known form in terms of modified Bessel functions~\cite{Schnedermann:1993ws,Florkowski:2010zz}
\begin{align}\label{eq:thkernel}
\begin{split}
K^\text{eq}_1(\pT ,  u^r)& = 4\pi m_\text{T} I_0\left(\frac{\pT  u^r}{T}\right)K_1\left(\frac{m_\text{T} u^\tau}{T}\right)\\
K^\text{eq}_2(\pT , u^r)& = 4\pi \pT  I_1\left(\frac{\pT  u^r}{T}\right)K_0\left(\frac{m_\text{T} u^\tau}{T}\right),
\end{split}
\end{align}
where $m_\text{T}=\sqrt{m^2+\pT ^2}$ and $T$ is the (constant) freeze-out temperature.
The 4-velocity $u^r = \beta/\sqrt{1-\beta^2}$ is given in terms of radial velocity $\beta(r)$, which is commonly parametrized by a power function with two parameters $\beta_\text{max}$ and $n$
\begin{equation}
    \beta(r) = \beta_\text{max}\frac{r^n}{r_\text{max}^n}.\label{eq:beta}
\end{equation}

\begin{figure}
\centering
\includegraphics[scale=0.35]{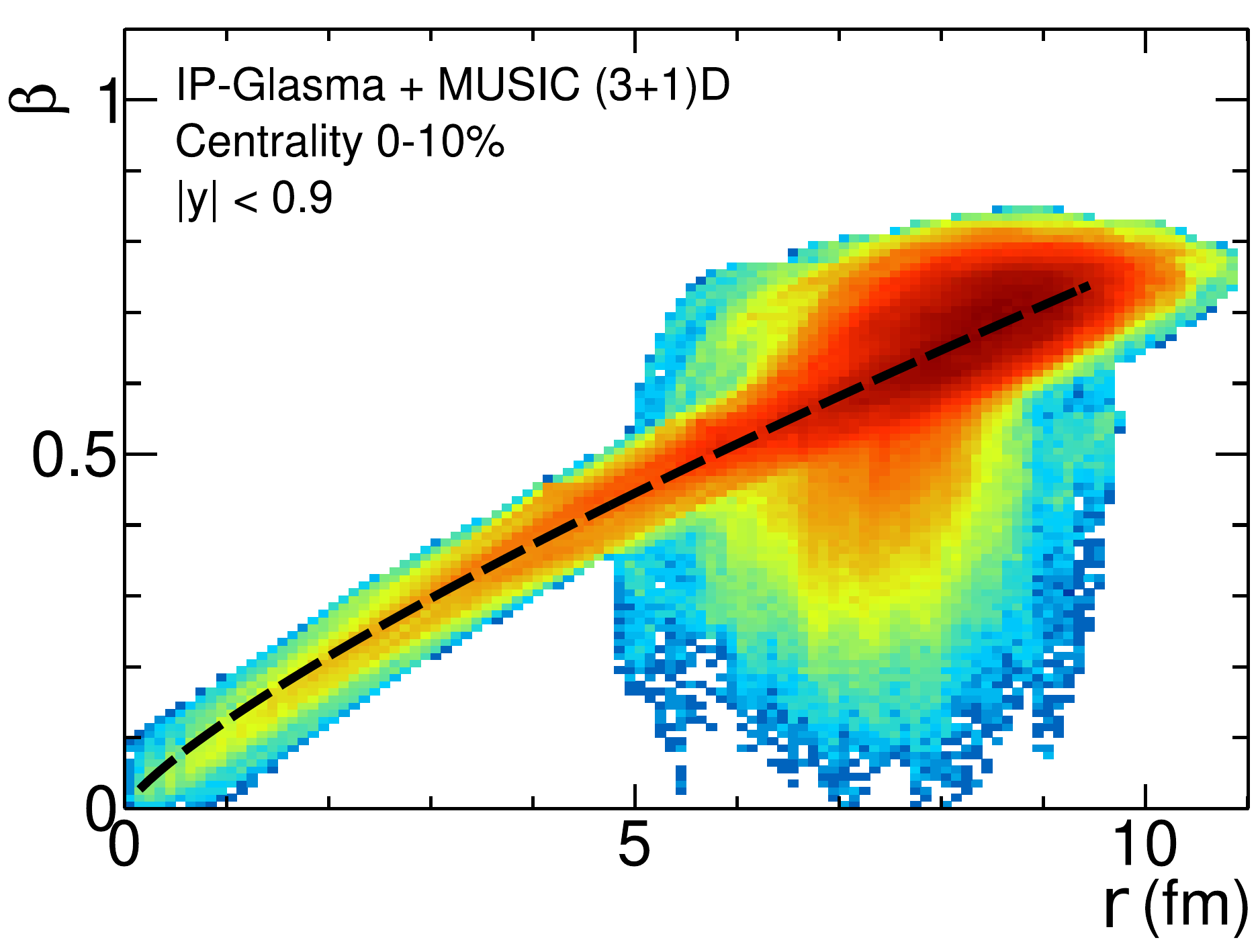}
\caption{Radial velocity profile on the freeze-out surface extracted from hydrodynamic simulations of central Pb-Pb collisions.}
\label{fig:plotv}
\end{figure}

In this paper the spectra of charmed hadrons formed in the core, i.e. by hadronization of the hot QGP fireball, are evaluated by using the velocity profile from a (3+1)D viscous hydrodynamics code MUSIC with IP-Glasma initial conditions tuned to the light flavor hadron observables~\cite{Schenke:2010nt,Schenke:2012wb}. The velocity profile and best fit with $\beta_\text{max}=0.62$ and $n=0.85$ for 0-10\% centrality bin is  shown in  Fig.~\ref{fig:plotv} (we use $\beta_\text{max}=0.60$ and $n=0.85$ for 30-50\% centrality bin). The fit uncertainties of the parameters $\beta_{\text{max}}$  and $n$ are 0.005 and 0.05, respectively.

\begin{figure}
    \centering
\includegraphics[width=0.49\linewidth]{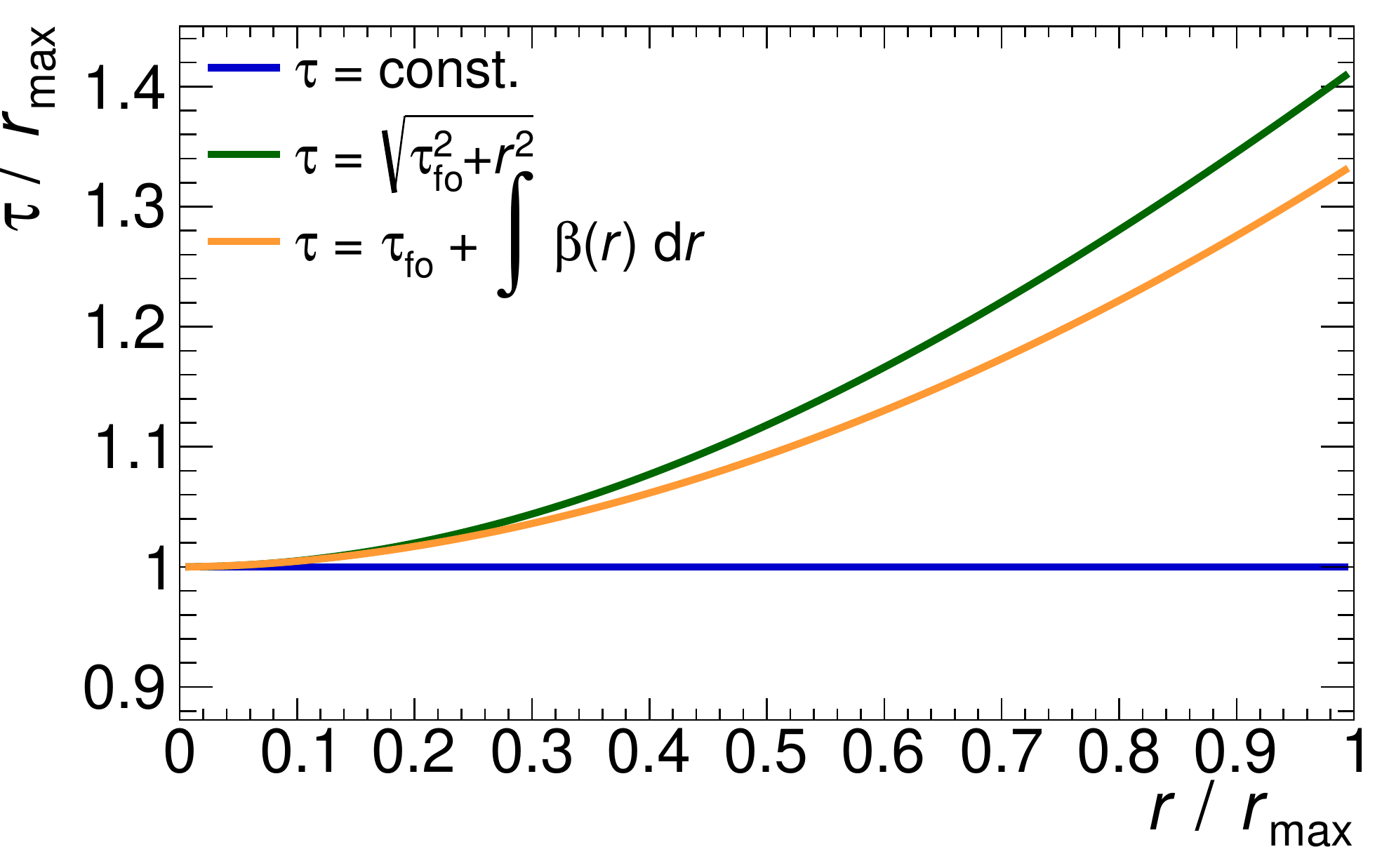}
\includegraphics[width=0.49\linewidth]{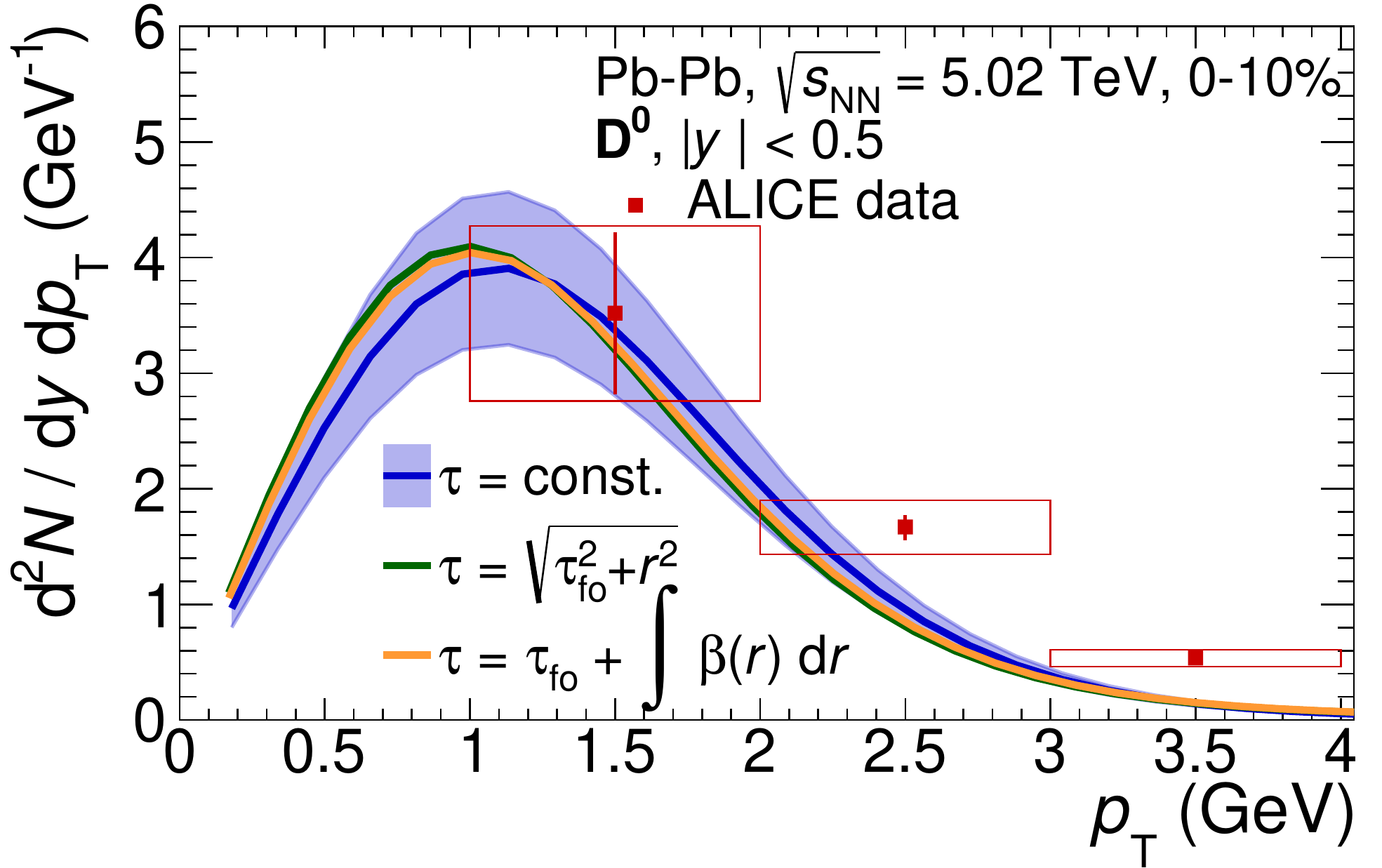}
    \caption{Left: freeze-out surface comparison, where $\tau_\text{fo}=r_\text{rmax}$. Right: $D^0$ spectra for different freeze-out surfaces. The shaded band is due to the normalization uncertainty in $g_c$. Experimentally measured points and their uncertainties~\cite{Acharya:2018hre} are shown for reference. }
    \label{fig:contour}
\end{figure}

Different types of freeze-out surfaces have been used in the past, for example, the constant Bjorken time freeze-out surface introduced in ref.~\cite{Schnedermann:1993ws}
\begin{align}
\tau(r) = \tau_\text{fo}
\end{align}
or constant proper time time surface of~\cite{Broniowski:2001uk}
\begin{align}
  \tau(r)&=\sqrt{\tau_\text{fo}^2+r^2}.
\end{align}
In ref.~\cite{Broniowski:2001uk} the velocity flow was restricted to be a Hubble-like   $u^\mu=x^\mu/\tau_\text{fo}$ and parallel to the norm of the surface.
For parametrized velocity in Eq.~\ref{eq:beta}, $u^\mu$ is no longer proportional to $d\sigma^\mu$. However, one can consider a third type of the surface for which this condition is still true: $\tau(r) = \tau_\text{fo}+\int_0^r dr'\beta(r')$ and using  Eq.~\ref{eq:beta} we get
\begin{equation}
  \tau(r)=\tau_\text{fo} + \frac{r\beta(r)}{n+1}.
\end{equation}
The three freeze-out surfaces are depicted in Fig.~\ref{fig:contour} (left). Without loss of generality, the freeze-out time is taken to be equal to $\tau_\text{fo}=r_\text{max}$ and $r_\text{max}$ itself can be determined by requiring the freeze-out volume per unit rapidity
\begin{align}
  V &= 2\pi\int_0^{r_\text{max}}\! \ud r \;  r \tau(r)u^\tau\left[1 -  \beta(r)\frac{\partial \tau}{\partial r}\right]
\end{align}
to be equal to a given value, e.g. $V=4997\,\text{fm}^3$ in central Pb-Pb collisions.
Note, however, that the integration variable $r$ can be rescaled to $ x = r/r_\text{max}$ with the result  that $r_\text{max}^3$ appears as normalization in front of the integral. Since we replace the overall normalization by that obtained from the SHMc, knowledge of $r_\text{max}$ is not required, and the only parameters left are the dimensionless parameters $\beta_{\text{max}}$ and $n$, as discussed above.

As we did in a previous publication for the J/$\psi$ spectrum~\cite{Andronic:2019wva}, the spectra for various charmed hadrons are computed using this velocity profile as input for a blast-wave parameterization in terms of temperature, flow velocity profile and mass of the hadron. The temperature we use  is the chemical freeze-out temperature $T_{cf} = 156.5\,\text{MeV}$  obtained from fitting the yields of light flavor hadrons and nuclei as measured by ALICE for Pb-Pb collisions at $\sqrt{s}$ = 2.76 TeV~\cite{Andronic:2017pug,Andronic:2018qqt}.
We studied the effects of the uncertainties of the blast wave parameters $\beta_{\text{max}}$  and $n$ on the hadron spectra. The resulting variations in the spectra are less than 10\% and in the ratios to $D^0$ less than 3\%.

In Fig.~\ref{fig:contour} (right) we show the $D^{0}$ spectra for the three freeze-out surfaces. We see that the difference in the absolute spectra is small and lies within the uncertainty band, which is mostly due to uncertainty in $g_c$ at these low momenta. In addition, given the still large experimental uncertainties we do not expect the precise form of the freeze-out surface to be the most important factor and we will use a constant freeze-out time surface as the default choice. We emphasize here that for particle ratios, e.g. $\LC/D^0$, even this small difference mostly cancels.

\begin{figure}
\centering
\includegraphics[width=.49\linewidth]{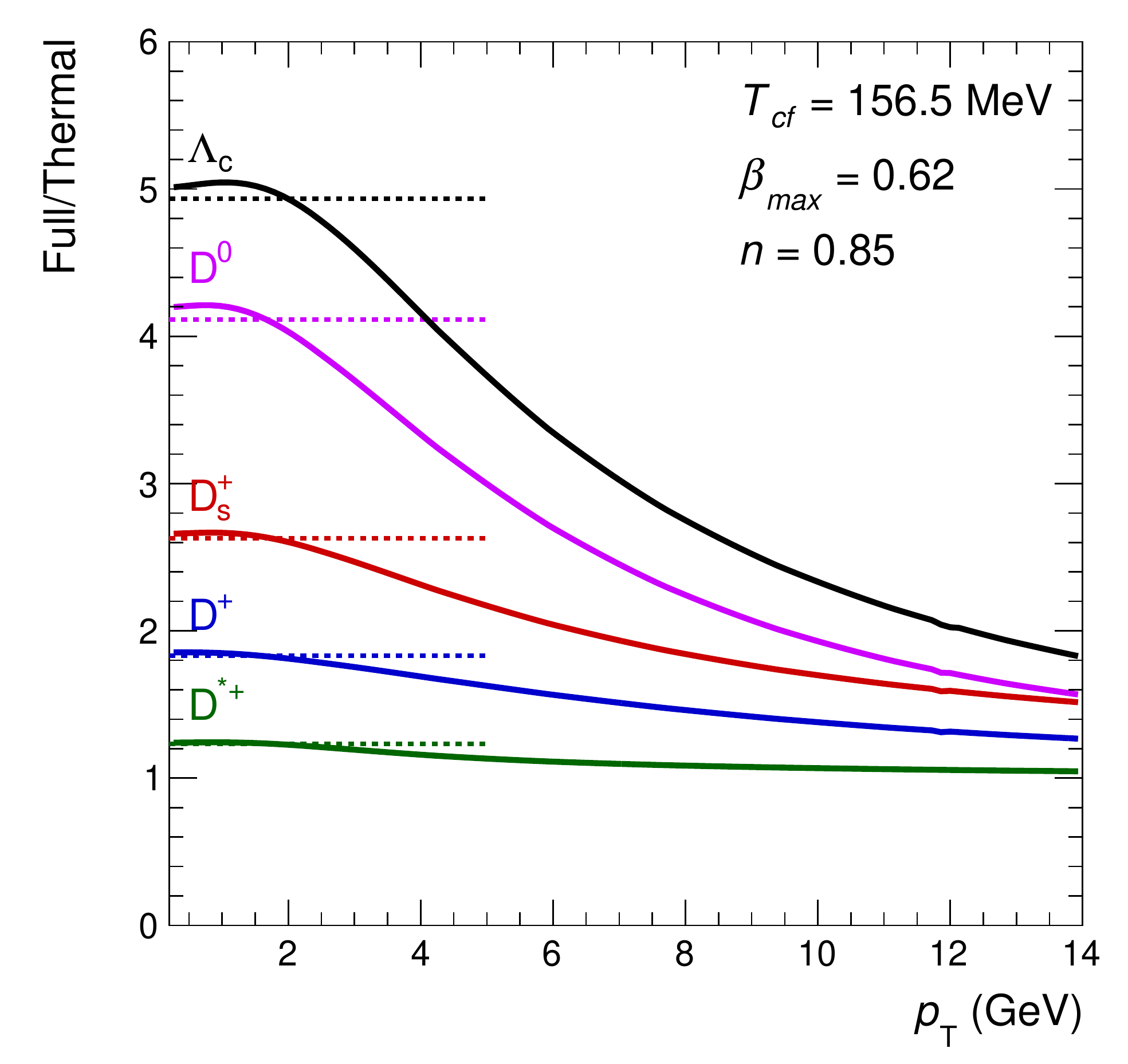}
\includegraphics[width=.49\linewidth]{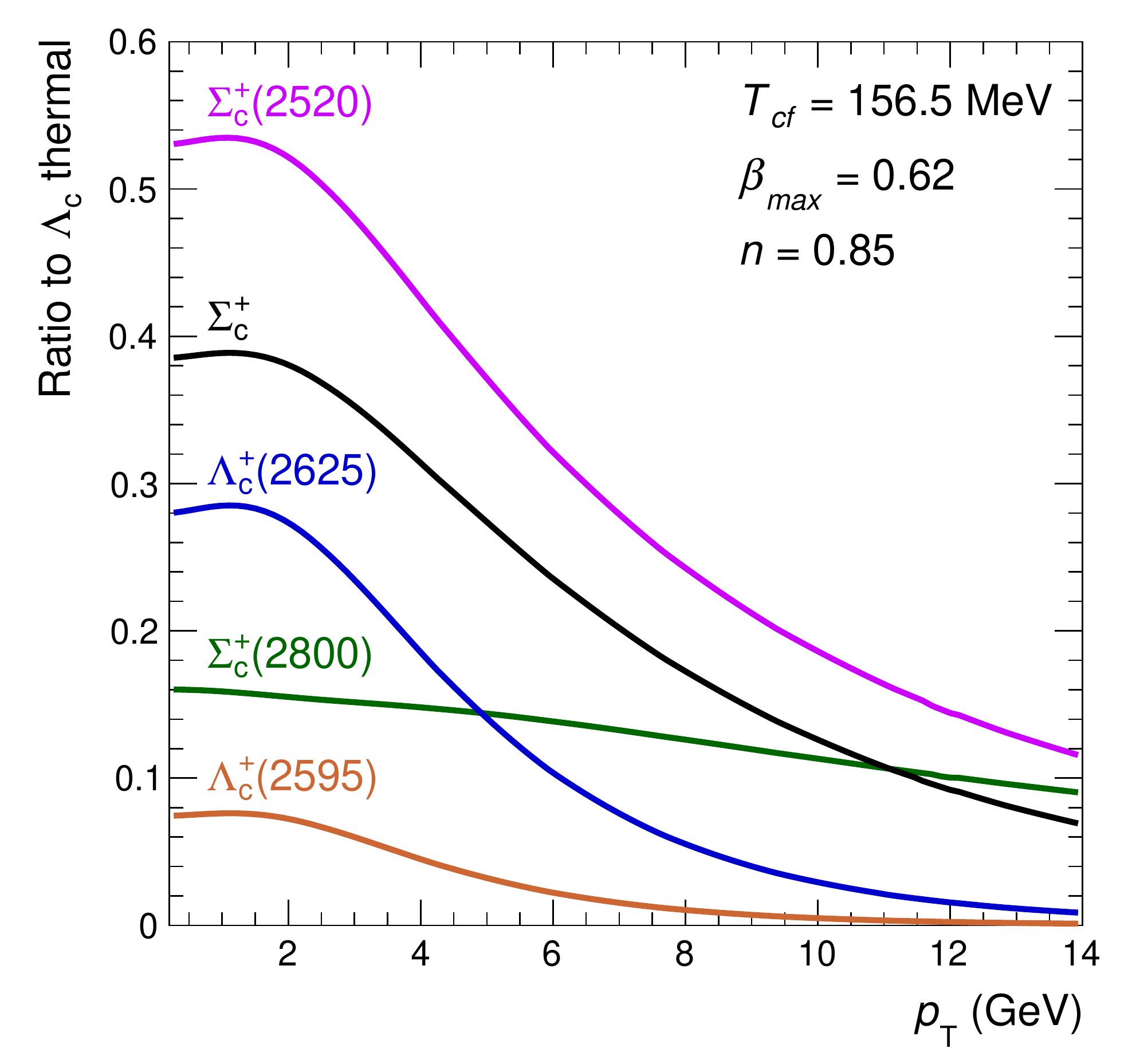}
\caption{Left: ratios of different particle spectra with feed-down contribution to thermal spectra (note that the corona contribution is not included here). Dashed lines correspond to the ratio of integrated yields (these ratios were previously used to scale thermal spectra in SHMc). Right: feed-down contribution to $\Lambda_c^+$ from different decay channels. For details see text.}
\label{fig:FdCorrection}
\end{figure}

One of the limitations of the standard blast-wave model is that it does not include the momentum modification of particle spectra due to the feed-down caused by resonance decays. Recently, a very efficient way of computing such modifications was derived~\cite{Mazeliauskas:2018irt} and applied in blast-wave fits with resonance decay feed-down~\cite{Mazeliauskas:2019ifr} and hydrodynamic simulations~\cite{Devetak:2019lsk}. Here we compute  the momentum resolved decay feed-down to long lived charmed mesons and baryons using the \texttt{FastReso} computer code~\cite{FastReso}.
In total we perform calculations for 76 $2$-body and 10 $3$-body decays of charmed mesons and baryons.
In practise, this procedure replaces thermal Boltzmann freeze-out kernels in Eq.~\ref{eq:Cooper-Frye} with numerically computed total final particle kernels.
We use the same temperature and radial velocity profiles as in a standard blast-wave model.  In Fig.~\ref{fig:FdCorrection} (left) we show the full decay spectra of charmed hadrons over their initial thermal spectra.
In addition, in Fig.~\ref{fig:FdCorrection}~(right)  we show the selected decay-channel contributions to $\LC$ spectra.
The feed-down contributions preferentially accumulate at low momentum and can be as large as  5 times that of thermal spectra for $\LC$. The dotted lines in Fig.~\ref{fig:FdCorrection} (left) show the ratio of full over thermal $\pT $-integrated yields in SHMc. These feed-down factors
were used previously to scale the thermal spectra without accounting for $\pT $ dependence of the feed-down. One can see rather good agreement between the naive and exact scaling of the spectra for $\pT \lesssim 3\,\text{GeV}$, where most of the particles are. As low momentum is the only region where core charmed hadron production is dominant, we find in practice very small differences between
full decay spectra and scaled thermal spectra in this momentum range. Nevertheless, in the plots below we will use the  spectra obtained with decay kernels from \texttt{FastReso}.

Finally, the high momentum power-law tail actually observed in experimental particle spectra is not described  by hydrodynamics.
Instead it can be modelled using a core-corona picture~\cite{Andronic:2019wva}.
Even in nucleus-nucleus collisions at small impact parameter, a number of nucleon-nucleon collisions take place in the so-called corona-region where the overlap density is a small fraction  of the maximum density achieved in the collision. In this overlap volume where nucleons undergo on average one or less collisions, we assume that no QGP is formed and, hence, treat the collisions as $\pp$-like. On the contrary, in the core part, we assume full thermalization of produced charm quarks.
We define the corona region as having 10\% of the central
density $\rho_0$. In a heavy nucleus at rest the central nucleon number density is $\rho_0 = 0.16\,{\rm fm}^{-3}$.
The \pT shape of the cross section measured in $\pp$ collision is parametrized  by
\begin{equation}
\frac{\ud^2\sigma^{\pp}}{\ud y \ud\pT }  = C \times \frac{\pT}{(1+(\pT/p_0)^2)^n},
  \label{eq:ppFit}
\end{equation}
where the coefficients $C$, $p_0$ and $n$ are obtained from a fit to experimental distributions for each particle species~\cite{Acharya:2021cqv, Acharya:2019mgn, Acharya:2020lrg} and the total integral of the function is set to  experimentally measured integrated cross section $\ud\sigma/\ud y$. The fit is found to describe the measured cross sections well within the uncertainties in the whole \pt range considered.  We then scale the $\pp$ differential cross section by the overlap function $T_\text{AA}^\text{corona}$ to account for the number of binary nucleon-nucleon collisions in the corona.

In summary, for each of the charmed hadrons under consideration the \pT{}  spectra are obtained by summing the soft momentum spectrum  from the the blast-wave model with resonance decays and the high momentum tail from the corona part. The uncertainty bands are obtained by varying $g_c$. In addition, the uncertainty on the corona part also includes the uncertainty of the fit to the pp data~\cite{Acharya:2021cqv,Acharya:2019mgn,Acharya:2020lrg}. This uncertainty is assumed to be uncorrelated for different particle species and is the dominant source of uncertainties for particle spectra and their ratios at high \pt, although it cancels for $R_\text{AA}$.

\section{Results for Pb-Pb and lighter collision systems}
\label{sec:results1}
\begin{figure*}
\centering
    \includegraphics[width=.49\linewidth]{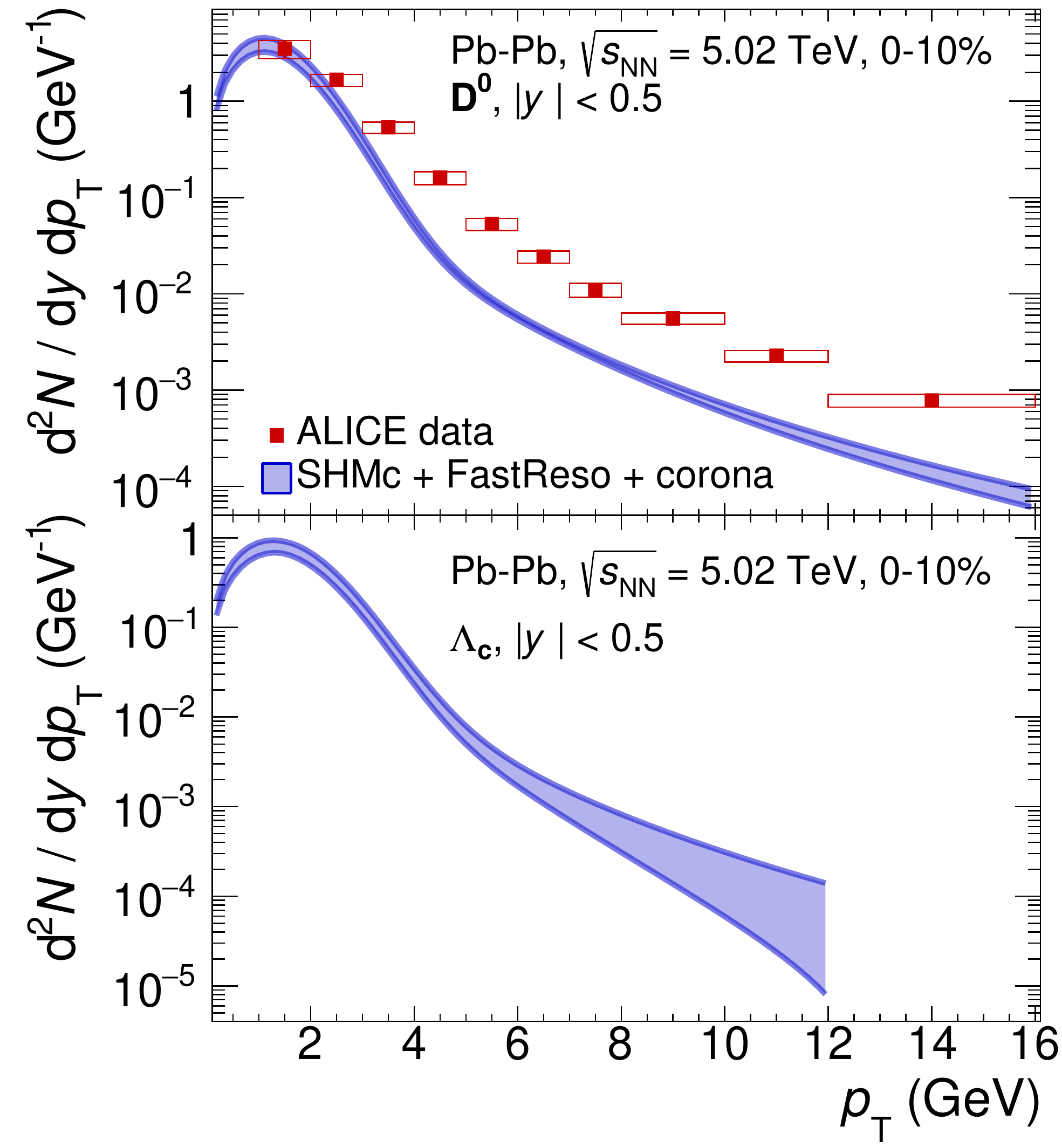}
    \includegraphics[width=.49\linewidth]{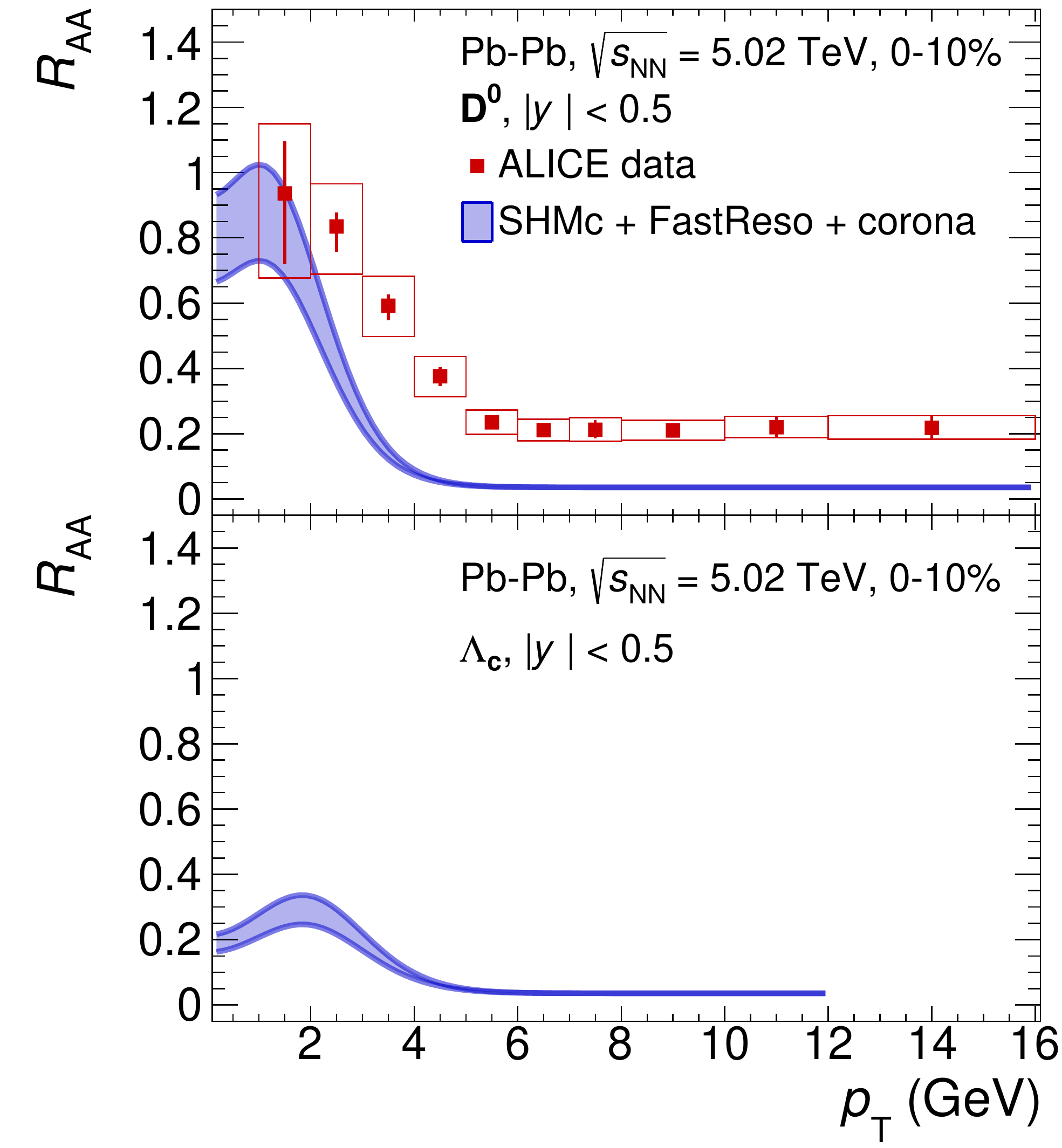}
    \caption{Spectra (left) and $R_{\rm AA}$ (right) of $\Dzero$ mesons (top) and $\Lambda_{\rm c}$ baryons (bottom) in Pb-Pb collisions at \cme{5.02} and 0-10\% centrality. Pb-Pb data for D-meson distributions taken from~\cite{Acharya:2018hre}. The pp data needed to compute the corona part are taken from~\cite{Acharya:2021cqv,Acharya:2020lrg}. The model band width at low and high \pt are driven by the uncertainties of $g_{c}$ and pp spectra fits, respectively, as described in the text.}
    \label{fig:spectra_1}
\end{figure*}

In the following we will describe predictions from the SHMc as well as the comparison of results from SHMc with the currently available data. For simplicity, we will only consider Pb-Pb collisions at \cme{5.02} and 0-10\% centrality, and predictions for 30-50\% centrality will be given in Appendix~\ref{sec:SemiCentralPredictions}. The model predictions for all particle species and the two centrality bins are available in numerical form as auxiliary file with the arXiv version of the publication.
By far the best series of experiments exists for $D$ mesons produced in Pb-Pb collisions, see~\cite{Acharya:2018hre}.

\subsection{Transverse momentum distributions}
In Fig.~\ref{fig:spectra_1} we show the comparison between the SHMc predictions and data for spectra and nuclear modification factor $R_{AA}$ as a function of transverse momentum $p_{\rm T}$. The transverse momentum dependence is obtained as explained in detail in section~\ref{sec:SHMc_pt} above.

Note that there are no new parameters used here apart from the hydrodynamics input discussed in section~\ref{sec:SHMc_pt}. The transverse momentum spectrum for $D^0$ mesons is very well described in particular in the purely thermal (``core") region for $p_{\rm T} \le 4$ GeV. In the transition region between core and corona as well for the high momentum tail we notice that the data are under-predicted for both the $p_{\rm T}$ spectrum and the $R_{AA}$. This suggests that the corona description is somewhat schematic and could be further optimized. The corresponding distribution for the $\Lambda_c$ baryon are displayed in the lower panels of Fig.~\ref{fig:spectra_1}. We note that these spectra and distributions are obtained with the unmodified charm resonance spectrum discussed below.
\begin{figure*}
\centering
\includegraphics[width=0.9\textwidth]{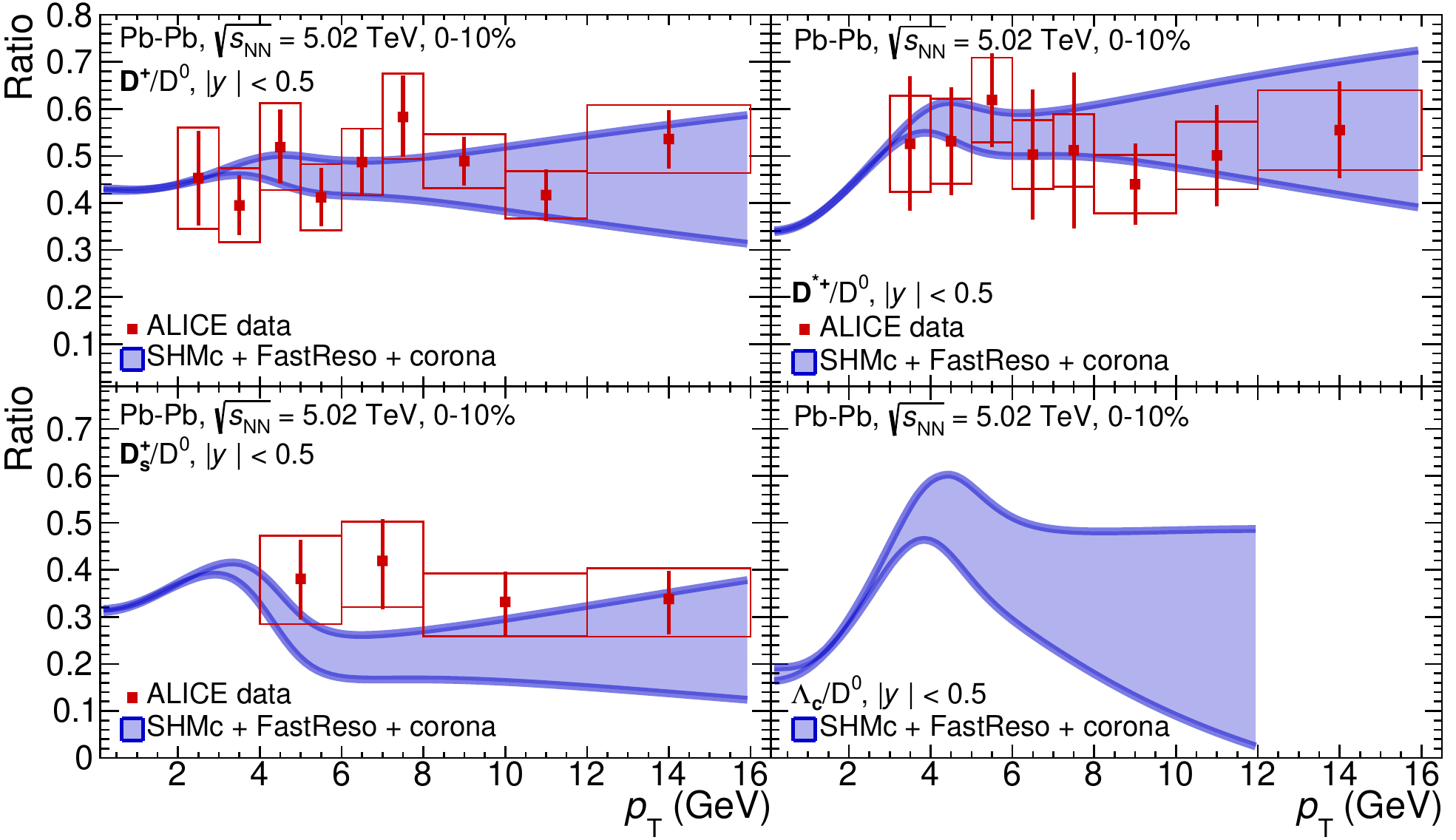}
\caption{Ratio of charmed hadron spectra, normalized to the $D^0$ spectrum from SHMc + FastReso + corona in Pb-Pb collisions at \cme{5.02} and 0-10\% centrality, in comparison to ALICE data~\cite{Acharya:2018hre}. The pp data needed to compute the corona part are taken from~\cite{Acharya:2021cqv,Acharya:2019mgn,Acharya:2020lrg}. The model band width at low and high \pt are driven by the uncertainties of $g_{c}$ and pp spectra fits, respectively, as described in the text.}
\label{fig:spectra_2}
\end{figure*}

In Fig.~\ref{fig:spectra_2} we show the corresponding distributions for $D^+$, $D^{*+}$, $D^{+}_{s}$ and $\Lambda_c$, plotted as a ratio to the $D^0$ spectrum. In this normalized plot, the charm cross section which determines the charm fugacity parameter $g_c$, is eliminated. For the three D-mesons we observe very good agreement  with the experimental observations. For the $\Lambda_c$ baryon the structure of the distribution changes quite strongly: a clear maximum appears near  $p_{\rm T} = 4.5$ GeV. Within the framework of the SHMc this maximum appears as a consequence of a superposition of collective flow (hydrodynamic expansion) and change of hadronization regime from bulk (statistical hadronization) to jets, much as it is observed also for the $\Lambda/K$ ratio in the (u,d,s) sector~\cite{Abelev:2013xaa}.

\subsection{Integrated yields}
In this section we discuss results for momentum integrated particle yields, which for constant temperature freeze-out assumed in the SHMc, do not depend on the details of the freeze-out surface and velocity prametrizations discussed in section~\ref{sec:SHMc_pt}

\begin{figure}
\centering
\includegraphics[width=0.49\textwidth]{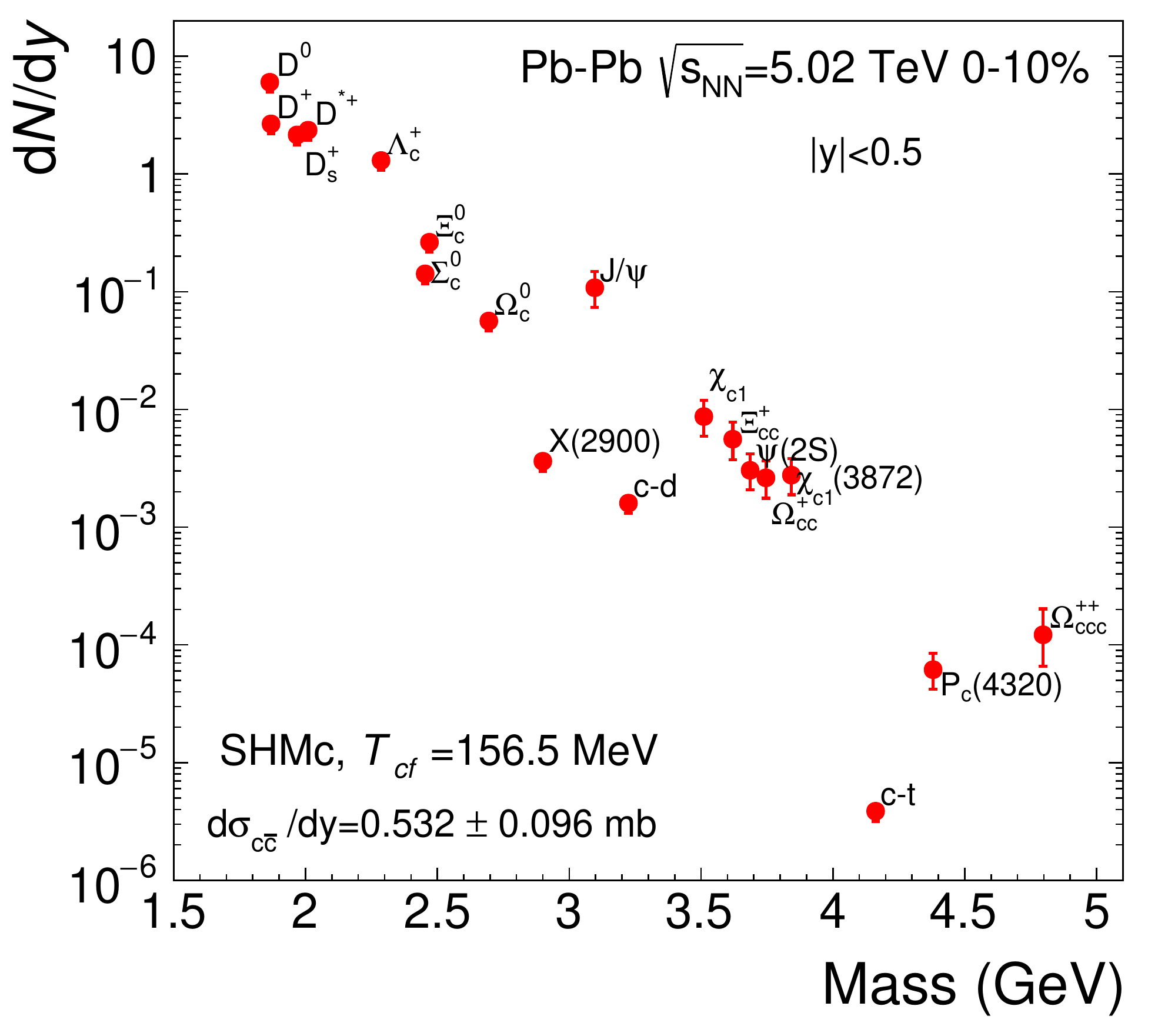}
\includegraphics[width=0.49\textwidth]{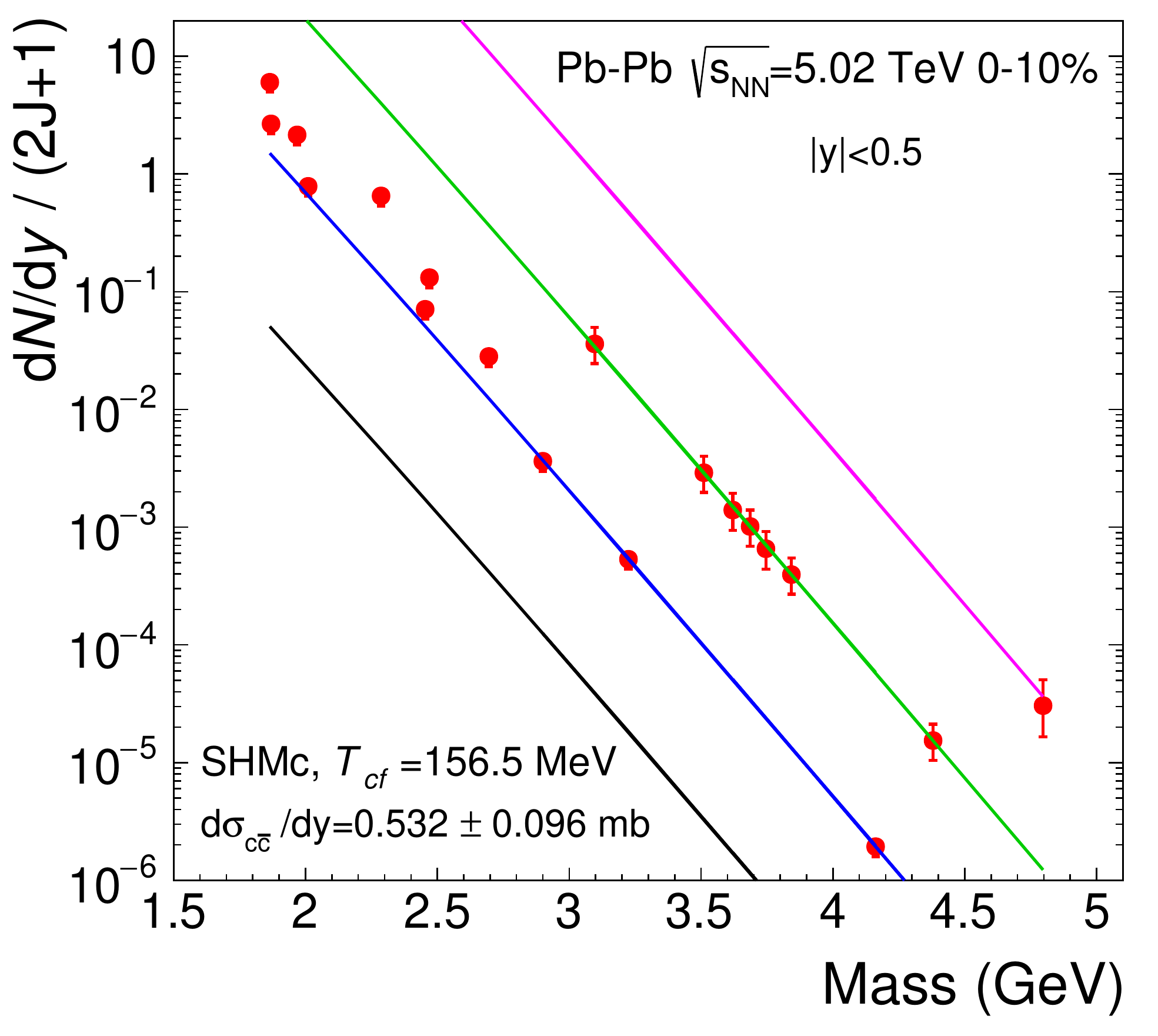}
\vskip -0.4 cm
\caption{Mass dependence of yields \dNdy~ for various hadron species for Pb-Pb collisions at mid-rapidity. The left panel is for absolute yields and the right panel is for yields per degree of freedom ($2J+1$). In this plot also the primordial (prior to decays) values are shown as lines, corresponding to hadrons with charm-quark or anti-quark content of 0, 1, 2, and 3 (respective powers of $g_c$).}
\label{fig:yields_m}
\end{figure}

\begin{figure}
\centering
\includegraphics[scale=0.4]{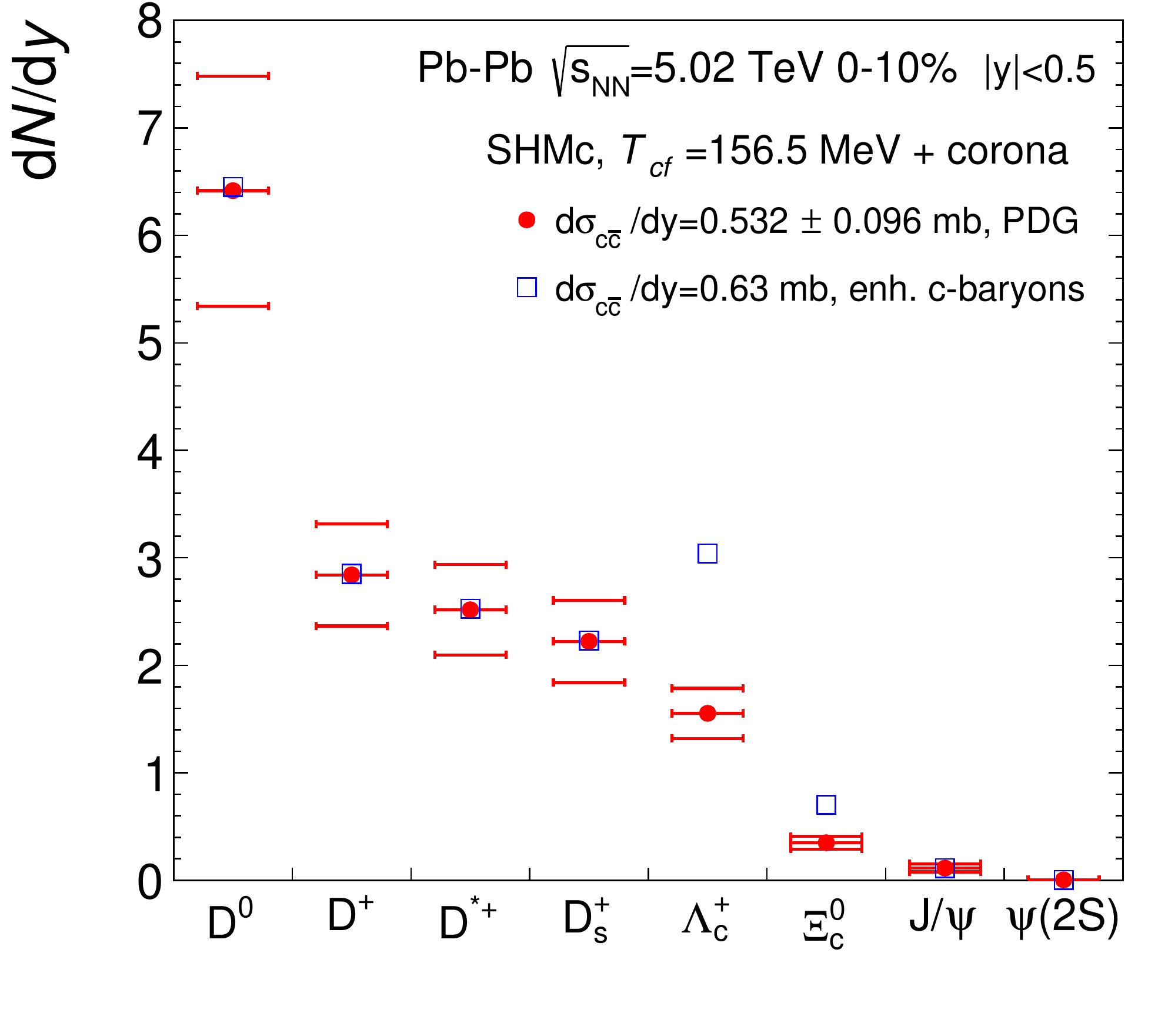}
\vskip -0.4 cm
\caption{Total (core+corona) yields \dNdy~ for various hadron species for central (0-10\%) Pb-Pb collisions at mid-rapidity. Red points correspond to the standard mass spectrum and total open charm cross section as discussed in the text. The open points where obtained with an enhanced total open charm cross section, implemented via tripled statistical weights for excited charmed baryons. For more details see text.}
\label{fig:yields_tot}
\end{figure}

  \begin{table*}
    \begin{tabular}{l | l l l }
Particle             &  $\ud N/\ud y$ core (SHMc)            &     \, $\ud N/\ud y$ corona              &   $\ud N/\ud y$ total                \\ \hline   \hline
 & \multicolumn{3}{c}{0-10\%}  \\
 \hline
 $D^{0}$             &   6.02 $\pm$ 1.07                     &    \,  0.396 $\pm$ 0.032                &    6.42 $\pm$ 1.07                    \\
 $D^{+}$             &   2.67 $\pm$ 0.47                     &    \,  0.175 $\pm$ 0.026                &    2.84 $\pm$ 0.47                    \\
 $D^{*+}$            &    2.36 $\pm$ 0.42                    &    \,  0.160 +0.048$-$0.022             &    2.52 $\pm$ 0.42                    \\
 $D_{s}^{+}$         &    2.15 $\pm$ 0.38                    &    \,   0.074 +0.024$-$0.015            &    2.22 $\pm$ 0.38                     \\
 $\Lambda_{c}^{+}$   &    1.30 $\pm$ 0.23                    &    \,   0.250 $\pm$ 0.028               &    1.55 $\pm$ 0.23                     \\
 $\Xi_{c}^{0}$       &    0.263 $\pm$ 0.047                  &    \,   0.090 $\pm$ 0.035               &    0.353 $\pm$ 0.058                    \\
 J/$\psi$           &     0.108 +0.041$-$0.035               &   \,  (5.08$\pm$0.37)$\cdot$10$^{-3}$   &    0.113 +0.041$-$0.035                \\
 $\psi(2S)$         &     (3.04 +1.2$-$1.0)$\cdot$10$^{-3}$  &  \,  (7.61$\pm$0.55)$\cdot$10$^{-4}$   &    (3.80 +1.2$-$1.0)$\cdot$10$^{-3}$    \\  \hline
  & \multicolumn{3}{c}{30-50\%}  \\
 \hline

 $D^{0}$             &    0.857 $\pm$ 0.153                    &  \,  0.207 $\pm$ 0.017               &   1.06 $\pm$ 0.154                    \\
 $D^{+}$             &    0.379 $\pm$ 0.068                    &  \,  0.092 $\pm$ 0.014               &   0.471 $\pm$ 0.069                   \\
 $D^{*+}$             &   0.335 $\pm$ 0.060                    &  \,  0.084 +0.025$-$0.011            &   0.419 +0.065$-$0.061                \\
 $D_{s}^{+}$          &   0.306 $\pm$ 0.055                    &  \,  0.039 +0.013$-$0.008            &    0.344 $\pm$ 0.056                   \\
 $\Lambda_{c}^{+}$    &    0.185 $\pm$ 0.033                   &  \,  0.131 $\pm$ 0.015               &    0.316 $\pm$ 0.036                   \\
 $\Xi_{c}^{0}$         &   0.038 $\pm$ 0.007                   &  \,  0.047 $\pm$ 0.018               &    0.084 $\pm$ 0.020                    \\
 J/$\psi$             &   (1.12 +0.37$-$0.32)$\cdot$10$^{-2}$ &  \, (2.65$\pm$0.19)$\cdot$10$^{-3}$   &   (1.39 +0.37$-$0.32)$\cdot$10$^{-2}$   \\
 $\psi(2S)$           &   (3.16 +1.04$-$0.89)$\cdot$10$^{-4}$ &  \, (3.98$\pm$0.29)$\cdot$10$^{-4}$   &   (7.14 +1.08$-$0.94)$\cdot$10$^{-4}$    \\

  \end{tabular}
    \caption{Summary of the calculations of yields at mid-rapidity for open charm and charmonia in Pb-Pb at 5.02 TeV, 0-10\% (upper part) and 30-50\% (lower part) centralities. For the corona, we used as inputs the production cross sections $\ud \sigma/\ud y$ as measured by ALICE in pp collisions \cite{Acharya:2019mgn,Acharya:2021cqv,Acharya:2020lrg,Acharya:2019lkw} (and assumed for $\Xi_c^0$ $\ud \sigma/\ud y$=0.10$\pm$0.04 mb and $\psi(2S)/\mathrm{J}/\psi = 0.15$) and $T_\text{AA}^\text{corona}$=0.90 mb$^{-1}$ and 0.47 mb$^{-1}$, respectively (for corona corresponding to $\rho<0.1\rho_0$). For details see text.} \label{tab:yields_tot}

  \end{table*}

In Fig.~\ref{fig:yields_m} we show the mass dependence of rapidity distributions  \dNdy~ for selected charm hadrons at mid-rapidity. The selection includes $D^0$ mesons at the lower masses and includes
many multi-charm states including the hypothetical $\Omega_{ccc}$ baryon at the high mass end of the plot. All are stable against decays via strong interactions.  Already the left plot exhibits clear structures whose origin becomes clear with the plot at the right hand side, where the yields are divided by the angular momentum degeneracy. Since we are in the 'Boltzmann' regime where all masses $M$ are much larger then the temperature $T_{cf} = 156.5$ MeV, the degeneracy-normalized particle yields scale in the SHMc as $\propto M^{3/2} \exp({-M/T_{cf}})$. In a log plot over 7 decades this function looks essentially like a straight line for fixed charm quark number. The color code separates particles with $\alpha = 1, 2, 3$ charm quarks. The line at the far left corresponds to $\alpha =0$ and coincides with that determined for (u,d,s) hadrons in~\cite{Andronic:2017pug}. The deviation clearly visible for $\alpha = 1$ is due to feeding from hadronically unstable resonances. The grouping into three distinct regions is what is called in the introduction 'the charm hadron hierarchy'.

In Fig.~\ref{fig:yields_tot} we show the total yields, the sum of core and corona components, for selected hadron species for which the data in pp collisions, used for the calculations of the corona component, are available. We include in the plot a scenario of charm baryon enhancement, implemented via tripled statistical weights for excited charmed baryons, which leads to an increase of the total thermal charm densities by 18\%. Note that the additional charmed baryon resonances are all assumed to be narrow Breit-Wigner-type resonances, as discussed in section~\ref{sec:SHMc_spec}. We demonstrate that the equivalent increase in the input charm cross section (from 0.53 to 0.63 mb) leads to a significant increase in the predicted yield for the charmed baryons, while the yields of all the rest of the species remain unchanged\footnote{After the completion of this work, the ALICE collaboration released~\cite{Acharya:2021set} a charm cross section at mid-rapidity for pp collisions at 5.02 TeV and based on the measurement of charmed mesons and baryons. Due to a significantly larger fragmentation into charmed baryons as compared to measurements in $\rm{e}^+\rm{e}^-$ and ep collisions, a charm cross section is obtained increased by 40\%  compared to the value on which the current calculations are based.}.
The numerical values for the case of the PDG hadron spectrum are shown in Table~\ref{tab:yields_tot}. One notices that some of the uncertainties are asymmetric and this originates either from SHMc, as the $g_c$ values are characterized by (slightly) asymmetric uncertainties and from the corona component via the experimental production cross section for pp collisions.

In Table~\ref{tab:yields_canonical12} we have compiled the expected luminosity, rapidity density for $\Omega_{ccc}$ production, inelastic cross section corresponding to the 10\% most central collisions, and expected yields for $\Omega_{ccc}$ production in 5 different collision systems at top LHC energy and for a run time of $10^6$ s. The beam parameters are from~\cite{Citron:2018lsq}, the rapidity densities and yields for $\Omega_{ccc}$ production are our predictions. The predictions are per unit rapidity for the 10\% most central collisions but contain no efficiency and acceptance corrections. Nevertheless, substantial yields can be expected. Even though the expected  luminosity increases by 4 orders of magnitude when moving from Pb-Pb to O-O, the yield in O-O is comparable to that for Pb-Pb, and that at a price of about 10 collisions per bunch crossing  for O-O~\cite{Citron:2018lsq}. Furthermore, corona effects will be much increased when going to such a small system. Which of the systems is optimal for QGP-related research will have to be carefully optimized.

\setlength{\tabcolsep}{4pt}
\begin{table*}
\begin{tabular}{l|ccccc}
 & O-O & Ar-Ar & Kr-Kr & Xe-Xe  & Pb-Pb\\
\hline\hline
 $\sigma_{\text{inel}}(10\%)\, \text{mb}$ & 140 & 260 & 420 &  580   & 800 \\
 $T_{\text{AA}}(0-10\%)\, \text{mb}^{-1}$ & 0.63 & 2.36 & 6.80 &  13.0   & 24.3 \\

$ \mathcal{L} ({\text{cm}^{-2}\text{s}^{-1}}) $   & $4.5 \cdot 10^{31}$ & $2.4 \cdot 10^{30} $ & $1.7\cdot 10^{29}$&  $3.0 \cdot 10^{28} $& $3.8 \cdot 10^{27}$ \\
\hline
&&&$\ud \sigma_{\ccBar}/\ud y = 0.53\,\text{mb}$  & &\\
\hline
$\ud N_{\Omega_{ccc}}/\ud y$   & $8.38 \cdot 10^{-8} $ & $1.29 \cdot 10^{-6} $ & $1.23 \cdot 10^{-5} $&  $4.17 \cdot 10^{-5}$  & $1.25 \cdot 10^{-4}$  \\
$\Omega_{ccc}$ Yield   & $5.3 \cdot 10^{5}$& $8.05 \cdot 10^5 $& $8.78 \cdot 10^5$  & $7.26 \cdot 10^5$ & $3.80 \cdot 10^5 $ \\
\hline
&&&$\ud \sigma_{\ccBar}/\ud y = 0.63\,\text{mb}$  & &\\
\hline
$\ud N_{\Omega_{ccc}}/\ud y$   & $1.44 \cdot 10^{-7} $ & $2.33 \cdot 10^{-6} $ & $2.14 \cdot 10^{-5} $&  $7.03 \cdot 10^{-5}$  & $2.07 \cdot 10^{-4}$  \\
$\Omega_{ccc}$ Yield   & $9.2 \cdot 10^{5}$& $1.45 \cdot 10^6 $& $1.53 \cdot 10^6$  & $1.22 \cdot 10^6$ & $6.29 \cdot 10^5 $
\end{tabular}
\caption{Expected yields for a run of $10^6$ s of $\Omega_{ccc}$ baryons for various collision systems at the LHC energy $\sqrt{s_{\mathrm{NN}}}=5.02$ TeV with full canonical suppression. All calculations are for mid-rapidity with $\Delta y = 1$. }
\label{tab:yields_canonical12}
\end{table*}

\section{Conclusions and Outlook}

In the present paper we have explored a range of predictions made within the framework of the SHMc with focus on hadrons with open charm. Most important is the comparison to recent ALICE measurements on $D$ mesons~\cite{Acharya:2018hre} and predictions for $\Lambda_c$ baryons. As baseline for SHMc predictions we kept the  chemical freeze-out temperature $T_{cf} = 156.5$ MeV determined from the analysis of (u,d,s) hadrons. As only additional input we used the open charm cross section based on pp measurements from the ALICE and LHCb collaborations and extrapolated to the Pb-Pb system using hard collision scaling and a correction for nuclear modifications obtained from an analysis of recently measured p-Pb open and hidden charm data. The transverse momentum distributions were obtained in a novel, hydro-inspired approach including resonance decays. Without any further assumptions and parameters all $D$ meson yields and low transverse momentum distributions in Pb-Pb collisions are well described. The situation is less well settled in the $\Lambda_c$ baryon sector. Recent ALICE measurements in pp and p-Pb collisions~\cite{Acharya:2020uqi} indicate enhanced production of $\Lambda_c$ baryons compared to what was expected based on $e^+e^-$  and $ep$ data on fragmentation into charmed baryons. For an account of ALICE preliminary data including those from Pb-Pb collisions see  Fig.~4 in~\cite{Loizides:2020tey}. These preliminary data have led to new charm baryon production models including ``missing" charm baryons~\cite{He:2019vgs}. We have therefore provided predictions for $\Lambda_c$ production in Pb-Pb collisions using the current experimental information on the charm baryon resonance spectrum~\cite{Zyla:2020zbs} as well as with an increased number of charm baryons. New data on this puzzling situation are expected soon from both the CERN ALICE and LHCb collaborations.

The success of the description of yields and low transverse momentum spectra of open charm hadrons by the SHMc also demonstrates that the hadronization of open and hidden charm takes place at or close to the QCD phase boundary. It further demonstrates that open and hidden charm data can be reproduced with one common hadronization mechanism.

Our predictions for Pb-Pb collisions imply very large enhancements for hadrons with 2 or 3 charm quarks compared to pure thermal production with charm fugacity $g_c =1$. The enhancement will be predominantly visible at low transverse momentum \pT{}, see, e.g., Fig.~\ref{fig:spectra_1}. For multi-charmed baryons these enhancements lead to an impressive and quite spectacular hierarchy, see Fig.~\ref{fig:yields_m}. To test these predictions is a challenge for future charm production experiments in LHC Run3 and Run4 and ultimately one of the important goals for the ALICE3 'all Silicon' experiment~\cite{Adamova:2019vkf}. Fundamental new information on the hadronization and deconfinement of charm quarks should be the rewards for the efforts to build such a detector.

\section{Acknowledgments}
\label{sec:Acknowledgments}
This work is part of and supported by the DFG (German Research Foundation) -- Project-ID 273811115 -- SFB  1225 ISOQUANT.
K.R. acknowledges the support by the Polish National Science Center (NCN) under the Opus grant no. 2018/31/B\-/ST2/01663, and   the  Polish  Ministry  of  Science  and  Higher Education. V.V. is supported by a research grant (Grant No. 00025462) from
VILLUM FONDEN, the Danish National Research Foundation
(Danmarks Grundforskningsfond), and the Carlsberg
Foundation (Carlsbergfondet).

\bibliographystyle{utphys}
\bibliography{references}

\appendix
\section{SHMc + FastReso + corona predictions for Pb-Pb collisions at \cme{5.02} and 30-50\% centrality} \label{sec:SemiCentralPredictions}
\begin{figure*}[h]
\centering
    \includegraphics[width=.49\linewidth]{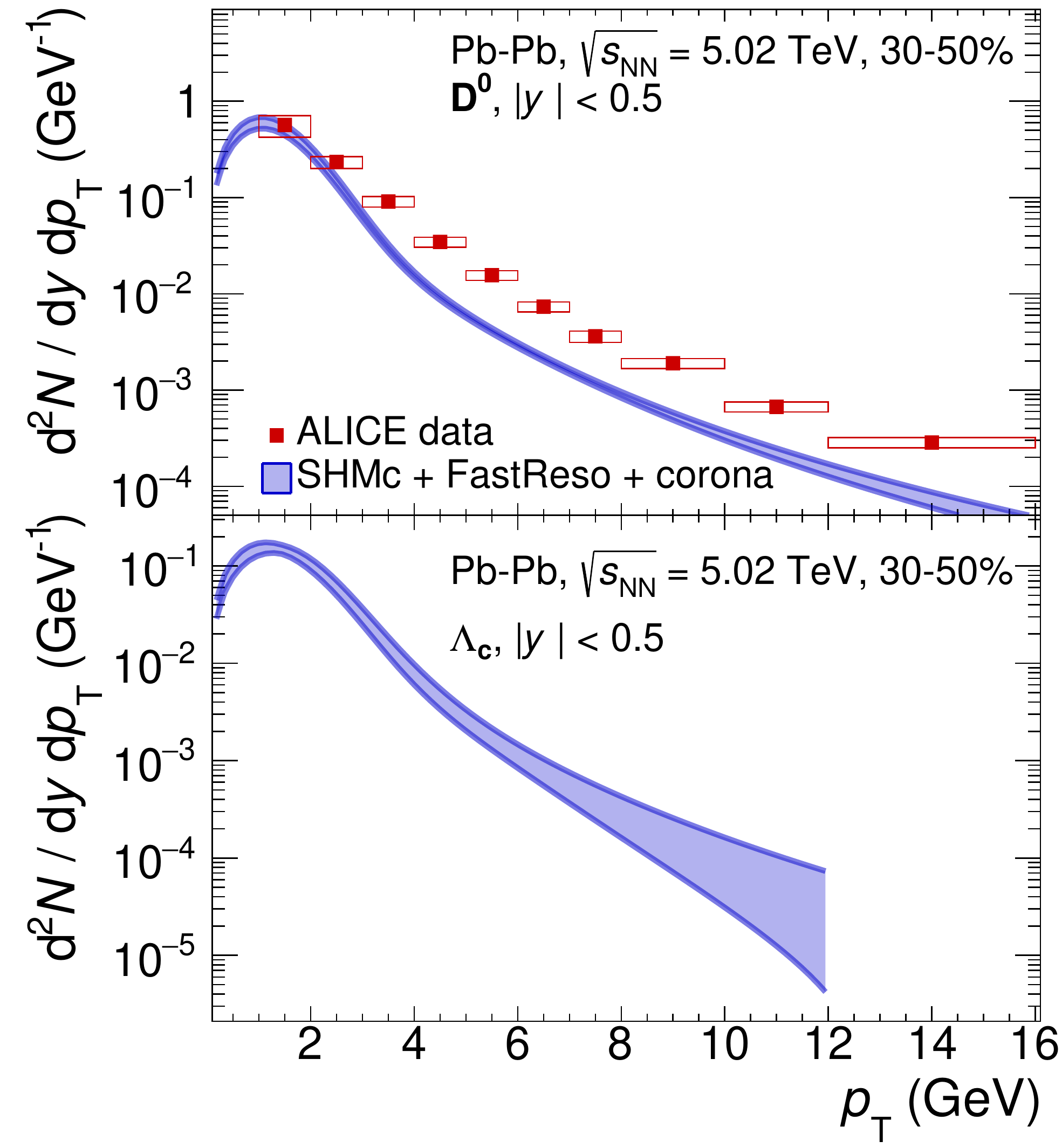}
    \includegraphics[width=.49\linewidth]{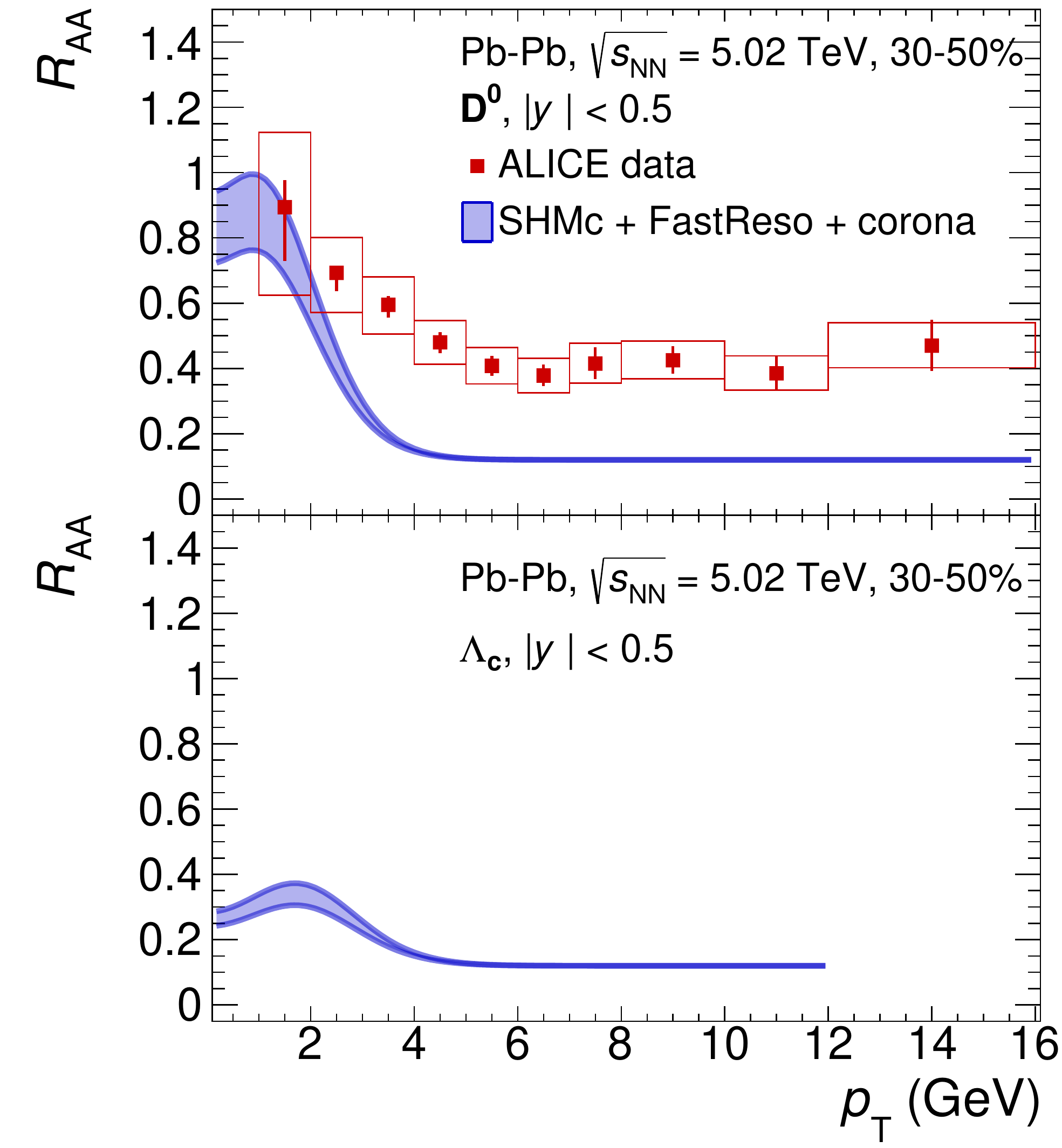}
    \caption{Spectra (left) and $R_{\mathrm{AA}}$ (right) of $\Dzero$ (top) and $\Lambda_{\rm c}$ in Pb-Pb collisions at \cme{5.02} and 30-50\% centrality. Pb-Pb data for D-meson distributions taken from~\cite{Acharya:2018hre}. The pp data needed to compute the corona part are taken from~\cite{Acharya:2021cqv,Acharya:2020lrg}. The model band width at low and high \pt are driven by the uncertainties of $g_{c}$ and pp spectra fits, respectively, as described in the text.}
    \label{fig:SemiCentral_1}
\end{figure*}

\begin{figure*}[h]
\centering
\includegraphics[width=0.9\textwidth]{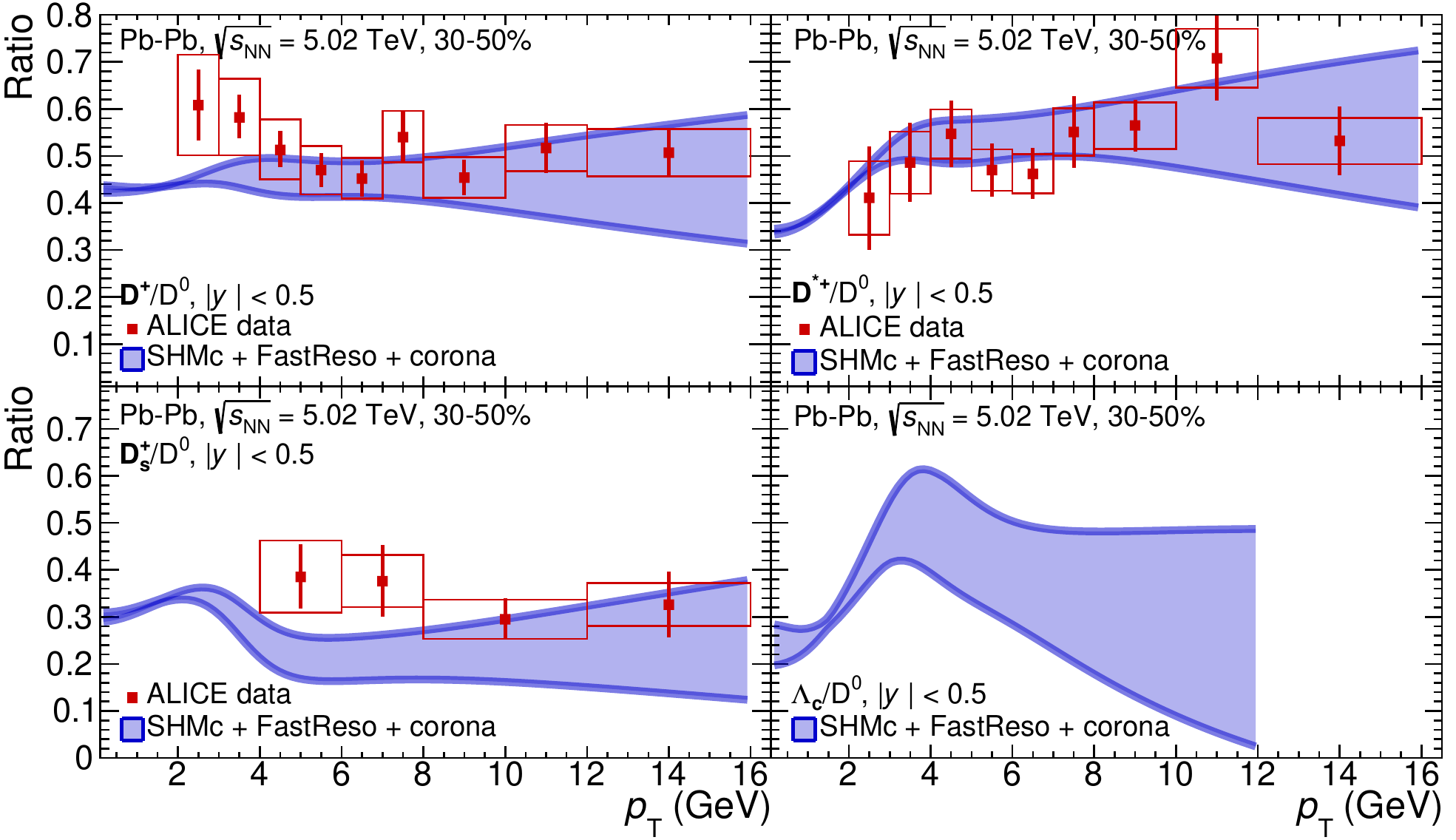}
\caption{Ratio of charmed hadron spectra, normalized to the $D^0$ spectrum from SHMc + FastReso + corona in Pb-Pb collisions at \cme{5.02} and 30-50\% centrality, in comparison to ALICE data~\cite{Acharya:2018hre}. The pp data needed to compute the corona part are taken from~\cite{Acharya:2021cqv,Acharya:2019mgn,Acharya:2020lrg}. The model band width at low and high \pt are driven by the uncertainties of $g_{c}$ and pp spectra fits, respectively, as described in the text.}
\label{fig:SemiCentral_2}
\end{figure*}

\end{document}